\documentclass[english,showpacs,prd,aps,superscriptaddress,preprintnumbers,notitlepage]{revtex4}
\usepackage[T1]{fontenc}
\usepackage[latin9]{inputenc}
\usepackage{babel}
\usepackage{amsmath}
\usepackage{amssymb}
\usepackage{graphicx}
\usepackage[unicode=true,
 bookmarks=false,
 breaklinks=false,pdfborder={0 0 1},backref=false,colorlinks=false]
 {hyperref}

\makeatletter

\usepackage{babel}
\PassOptionsToPackage{normalem}{ulem}
\usepackage{ulem}

\usepackage{epsfig}
\usepackage{amsfonts}

\makeatother

\begin{document}

\title{Multiplicity distributions as probes of quarkonia production mechanisms}

\author{Marat Siddikov}
\email{marat.siddikov@usm.cl}

\affiliation{Departemento de Física, Universidad Técnica Federico Santa María,
and Centro Científico-~~~~~\\
 Tecnológico de Valparaíso, Avda. España 1680, Casilla 110-V, Valparaíso,
Chile}

\author{Eugene Levin}
\email{leving@tauex.tau.ac.il, eugeny.levin@usm.cl}

\affiliation{Departemento de Física, Universidad Técnica Federico Santa María,
and Centro Científico-~~~~~\\
 Tecnológico de Valparaíso, Avda. España 1680, Casilla 110-V, Valparaíso,
Chile}

\affiliation{Department of Particle Physics, School of Physics and Astronomy,
Raymond and Beverly Sackler Faculty of Exact Science, Tel Aviv University,
Tel Aviv, 69978, Israel}

\author{Iv\'an Schmidt}
\email{ivan.schmidt@usm.cl}

\affiliation{Departemento de Física, Universidad Técnica Federico Santa María,
and Centro Científico-~~~~~\\
 Tecnológico de Valparaíso, Avda. España 1680, Casilla 110-V, Valparaíso,
Chile}

\date{\today}


\begin{abstract}
In this paper we demonstrate that the vigorously growing multiplicity
distributions measured by STAR and ALICE present a strong evidence
in favor of multigluon fusion mechanisms of the quarkonia production
in CGC approach. We analyze the contribution of 3-gluon fusion mechanism
and demonstrate that it gives a sizeable contribution to quarkonia
yields, as well as predicts correctly the multiplicity distributions
for $J/\psi$ at RHIC and LHC. We also make predictions for other
quarkonia states, such as $\psi(2S)$ and $\Upsilon(1S)$, and find
that the multiplicity dependence of these states should be comparable
to similar dependence for $J/\psi$. Finally, we discuss an experimental
setup in which very strong multiplicity dependence could be observed.
This observation would be a strong evidence in favor of CGC approach.
\end{abstract}

\preprint{TAUP-3039/19, USM-TH-365}
\maketitle

\section{Introduction}

The production mechanisms of hadrons which contain heavy quarks (open-charm
$D$-mesons and charmonia states) remains one of the long-standing
puzzles of high energy particles physics. The standard approaches
assume that the dominant contribution comes from gluon-gluon fusion,
and the heavy quarks formed in the process might emit soft gluons
in order to hadronize into $D$-mesons or form quarkonia states~\cite{Maciula:2013wg,Chang:1979nn,Baier:1981uk,Berger:1980ni}.
This picture gives reasonable estimates for the total and differential
cross-sections, within uncertainty due fragmentation functions of
$D$-mesons or Long Distance Matrix Elements (LDMEs) of charmonia
states~\cite{Bodwin:1994jh,Maltoni:1997pt,Brambilla:2008zg,Feng:2015cba,Brambilla:2010cs,Baranov:2015laa,Baranov:2016clx}.

However, the descriptions based on gluon-gluon fusion picture hardly
can explain the multiplicity distributions of heavy mesons recently
measured at LHC. As was found by ALICE~\cite{ALICE:AAStrangeness,ALICE:AAStrangeness2}
collaboration, the charmonia yields grow rapidly as a function of
the multiplicity of co-produced charged particles. Early results from
$AA$ collisions were interpreted as a telltale sign of Quark-Gluon
Plasma formation, and were successfully described in the framework
of the hydrodynamic models. However, recent measurements of ALICE
demonstrated that similar enhancement occurs in $pA$~\cite{ALICE:pAStrangeness,ALICE:pAStrangeness2}
and even in $pp$ collisions~\cite{ALICE:2017jyt,Thakur:2018dmp},
as well as in production of $D$-mesons~\cite{Adam:2015ota}. The
analysis of experimental data did not find a significant dependence
on either the initial volume (collision centrality) or the initial
energy density (which is correlated with collision energy). A similar
enhancement occurs in strangeness production, where the yield ratios
$\left(\Lambda+\bar{\Lambda}\right)/2K_{S}^{0}$ and $\left(p+\bar{p}\right)/\left(\pi^{+}+\pi^{-}\right)$
do not change significantly with multiplicity, thus establishing that
the observed effect is not due to the difference in the hadron masses.
As was discussed in~\cite{Fischer:2016zzs}, these new findings cannot
be easily accommodated in the framework of models based on gluon-gluon
fusion picture and potentially could require introduction of new mechanisms
both for $AA$ and $pp$ collisions.

\begin{figure}[ht]
\centering \leavevmode \includegraphics[width=14cm]{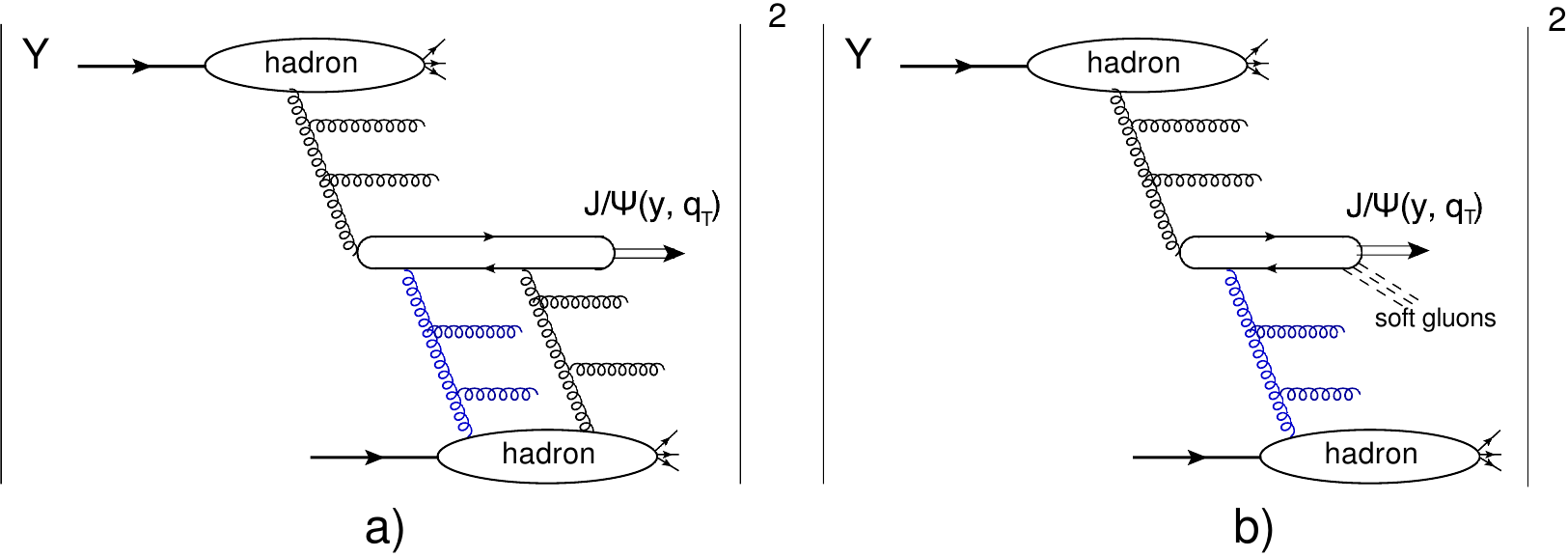}
\caption{Quarkonia production via three-pomeron fusion (a) and the two-pomeron
fusion with additional soft gluon emission (b). }
\label{2sh} 
\end{figure}

Recently in~\cite{LESI} we suggested that the experimentally observed
dependence on multiplicity in quarkonia production~\cite{PSIMULT,Alice:2012Mult}
could indicate a sizeable contribution of the multipomeron mechanisms~\footnote{In what follows we call ``pomeron'' a gluon cascade (``shower'')
which in the Figure~\ref{2sh} corresponds to the initial state \textquotedblleft cut
ladder'' and which leads to production of hadrons almost uniformly
distributed in the entire region of rapidities. Such parton showers
correspond to the cut BFKL Pomeron.}.
While it is expected that such higher-order contributions should be
suppressed in the heavy quark mass limit, for charmonia such suppression
does not work due to compensation of the large-$m_{c}$ suppression
by enhanced gluon densities in small-$x$ kinematics. Besides, for
the 3-pomeron term such suppression does not work even in the heavy
quark mass limit and therefore 3-pomeron term gives a sizeable contribution~\cite{LESI,KMRS,MOSA}.
Indeed, as we can see from the Fig. (\ref{2sh}), formally the two-pomeron
and three-pomeron mechanisms contribute in the same order in the strong
coupling $\alpha_{s}$. For a long time it was believed that two-pomeron
mechanism dominates, since the softness of the emitted gluons in the
two-pomeron mechanism implies milder suppression in the heavy quark
mass limit than expected from $\mathcal{O}(\alpha_{s})$ -counting.
However, at high energy the contribution of the three-pomeron mechanism
is also enhanced because the additional $t$-channel gluon yields
enhancement due to increased gluon densities.  As was illustrated
in~\cite{LESI,KMRS,MOSA}, the latter mechanism also provides a reasonable
description of the rapidity and the $p_{T}$-dependence of produced
charmonia~\footnote{In case of nuclear $pA$ and $AA$ collisions this mechanism gets
additional enhancement $\sim A^{1/3}$ and gets an additional enhancement
compared to other mechanisms, as was demonstrated in~\cite{KHTU}
for proton-nucleus and in~\cite{KLNT,DKLMT,KLTPSI,KMV,GOLEPSI} for
nucleus-nucleus scattering.}. 

In our preliminary study~\cite{LESI} we demonstrated that the three-pomeron
mechanism qualitatively can describe the main features of the multiplicity
distribution. In this paper we analyze the multiplicity dependence
in detail and apply the general CGC/saturation approach to its evaluation
in different kinematics and different quarkonia states.

The paper is structured as follows. In the next Section~\ref{sec:ProdMec}
we describe a framework for quarkonia production in the CGC/Saturation
approach. In Section~\ref{sec:Numer} we make numerical estimates
of the multiplicity distributions, compare with available experimental
data, and make predictions for the $\psi(2S)$ and $\Upsilon(1S)$
quarkonia. Finally, in Section~\ref{sec:Concusion} we draw conclusions
and discuss some open challenges in our approach.

\section{Three-pomeron contribution to charmonia production}

\label{sec:ProdMec}

\label{subsec:3Pom}The contribution of the three-pomeron mechanism
to the cross-section of the $S$-wave quarkonia production is given
by~\cite{LESI} 
\begin{eqnarray}
 &  & \frac{d\sigma\left(y,\,\sqrt{s}\right)}{dy}\,\,=\,\,\,2\,\int\frac{d^{2}Q_{T}}{(2\pi)^{2}}\,S_{h}^{2}\left(Q_{T}\right)\,\,x_{1}(y)G\left(x_{1}(y),\,\mu_{F}\right)\,\,\label{FD1}\\
 &  & \times\,\,\,\int_{0}^{1}dz\int_{0}^{1}dz'\int\frac{d^{2}r}{4\pi}\,\frac{d^{2}r'}{4\pi}\,\,\,\left\langle \Psi_{g}\left(r,\,z\right)\,\Psi_{M}\left(r,\,z\right)\right\rangle \,\left\langle \Psi_{g}\left(r',\,z'\right)\,\Psi_{M}\left(r',\,z'\right)\right\rangle \nonumber \\
 &  & \times\,\,\Bigg(\,N_{G}\left(x_{2}(y);\,\frac{\vec{r}+\vec{r}'}{2}\right)\,-\,N_{G}\left(x_{2}(y);\,\frac{\vec{r}-\vec{r}'}{2}\right)\Bigg)^{2}+\left(x_{1}\leftrightarrow x_{2}\right),\nonumber \\
 &  & x_{1,2}\approx\frac{\sqrt{m_{M}^{2}+\langle p_{\perp M}^{2}\rangle}}{\sqrt{s}}e^{\pm y}
\end{eqnarray}
where $y$ is the rapidity of produced quarkonia measured in the center-of-mass
frame of the colliding protons, $\Psi_{M}(r,\,z)$ is the light-cone
wave function of the quarkonium $M$ ($M=J/\psi,\,\psi(2S),\,\Upsilon(1S)$)
with transverse separation between quarks $r$ and the light-cone
fraction carried by the quark $z$; $\Psi_{g}$ is the wave function
of the heavy quark-antiquark pair. We also use notation $N_{G}=2N\,-\,N^{2}$
where $N\left(y,\,\vec{r}\right)\equiv\int d^{2}b\,N\left(y,r,b\right)$
is the dipole scattering amplitude. The factors in the first line
are explained in detail in~\cite{LESI} and are related to gluon
uPDFs and DPDFs in the proton. The nonperturbative factor $\sim\int d^{2}Q\,S_{h}^{2}\left(Q_{T}\right)$
is important for the absolute cross-sections studied in~\cite{LESI},
yet eventually will cancel in self-normalized observables considered
in this paper. The second factor includes the gluon PDF $\,x_{g}G\left(x_{g},\mu_{F}\right),$
which we take at the scale $\mu_{F}\,\approx2\,m_{c}$. In the LHC
kinematics at central rapidities (our principal interest) this scale
significantly exceeds the saturation scale $Q_{s}(x)$, which justifies
the use of two-gluon approximation. However, in the kinematics of
small-$x_{g}$ (large energies) there are sizeable non-perturbative
(nonlinear) corrections to evolution in the CGC/Saturation approach.
The corresponding scale $\mu_{F}$ in this kinematics should be taken
at the saturation momentum $Q_{s}$. The gluon PDF $x_{1}G\left(x_{1},\,\mu_{F}\right)$
in this approach is closely related to the dipole scattering amplitude
$N\left(y,r\right)=\int d^{2}b\,N\left(y,r,b\right)$ as~\cite{KOLEB,THOR}
\begin{equation}
\frac{C_{F}}{2\pi^{2}\bar{\alpha}_{S}}N\left(y,\,\vec{r}\right)=\int\frac{d^{2}k_{T}}{k_{T}^{4}}\phi\left(y,k_{T}\right)\,\Bigg(1-e^{i\vec{k}_{T}\cdot\vec{r}}\Bigg);~~~~x\,G\left(x,\,\mu_{F}\right)=\int_{0}^{\mu_{F}}\frac{d^{2}k_{T}}{k_{T}^{2}}\phi\left(x,\,k_{T}\right),\label{GN1}
\end{equation}
where $y=\ln(1/x)$. The Eq. (\ref{GN1}) might be inverted and gives
the gluon uPDF in terms of the dipole amplitude, 
\begin{equation}
xG\left(x,\,\mu_{F}\right)\,\,=\,\,\frac{C_{F}\mu_{F}}{2\pi^{2}\bar{\alpha}_{S}}\int d^{2}r\,\frac{J_{1}\left(r\,\mu_{F}\right)}{r}\nabla_{r}^{2}N\left(y,\,\vec{r}\right).\label{GN2}
\end{equation}

The result allows us to rewrite~(\ref{FD1}) in a more symmetric
form, entirely in terms of the dipole amplitude $N$,{\small{} 
\begin{eqnarray}
 &  & \frac{d\sigma\left(y,\,\sqrt{s}\right)}{dy}\,\,=\frac{2C_{F}\mu_{F}}{\bar{\alpha}_{S}\,\pi}\,\int\frac{d^{2}Q_{T}}{(2\pi)^{2}}\,S_{h}^{2}\left(Q_{T}\right)\,\,\times\label{FD3}\\
 &  & \times\,\,\int_{0}^{1}dz\int_{0}^{1}dz'\int\frac{d^{2}r}{4\pi}\,\frac{d^{2}r'}{4\pi}\,\,\,\left\langle \Psi_{g}\left(r,z\right)\,\Psi_{M}\left(r,z\right)\right\rangle \,\left\langle \Psi_{g}\left(r',z'\right)\,\Psi_{M}\left(r',z'\right)\right\rangle \nonumber \\
 &  & \times\,\,\int dr"J_{1}\left(\mu_{F}\,r''\right)\Bigg\{\nabla^{2}N\left(x_{1},\,r"\right)\Bigg(\,N_{G}\left(x_{2};\,\frac{\vec{r}+\vec{r}'}{2}\right)\,-\,N_{G}\left(x_{2};\,\frac{\vec{r}-\vec{r}'}{2}\right)\Bigg)^{2}\,\,\nonumber \\
 &  & +\,\,\,\nabla^{2}\,N\left(x_{2},\,r"\right)\,\Bigg(\,N_{G}\left(x_{1};\,\frac{\vec{r}+\vec{r}'}{2}\right)\,-\,N_{G}\left(x_{1};\,\frac{\vec{r}-\vec{r}'}{2}\right)\Bigg)^{2}\Bigg\}.\nonumber 
\end{eqnarray}
} which will be used for analysis below{\small{}. }{\small \par}

\section{Numerical estimates}

\label{sec:Numer}In CGC/saturation approach the dipole amplitude
$N\left(y,\,\vec{r},\,\vec{b}\right)$ is expected to satisfy the
non-linear Balitsky-Kovchegov\cite{BK} equation for the dipoles of
small size $r$. In the saturation region this solution should exhibit
a geometric scaling, being a function of one variable $\tau\,=\,r^{2}\,Q_{s}^{2}$,
where $Q_{s}$ is the saturation scale~\cite{GS1,GS2,GS3,LETU}.
Such behaviour is implemented in different phenomenological parametrizations
available from the literature. One of such parametrizations which
we will use for our numerical estimates is the CGC parametrization~\cite{RESH}
(for the sake of definiteness, we use the first set of parameters
from Table I in~\cite{RESH}).

As was illustrated in~\cite{LESI}, the quarkonia production cross-section~(\ref{FD3})
with CGC dipole parametrization provides a very reasonable description
of the shapes of produced charmonia as a function of rapidity and
transverse momentum, as well as predicts that the suggested mechanism
gives the dominant contribution to $J/\psi$ yields. The description
of multiplicity dependence presents more challenges at the conceptual
level because there are different mechanisms to produce enhanced number
of charged particles $N_{{\rm ch}}$. The probability of multiplicity
fluctuations decreases rapidly as a function of number of produced
charged particles $N_{{\rm ch}}$~\cite{Abelev:2012rz}, for this
reason for study of the multiplicity dependence it is more common
to use a normalized ratio~\cite{Thakur:2018dmp} 
\begin{align}
 & \frac{dN_{J/\psi}/dy}{\langle dN_{J/\psi}/dy\rangle}\,\,=\frac{w\left(N_{J/\psi}\right)}{\left\langle w\left(N_{J/\psi}\right)\right\rangle }\,\frac{\left\langle w\left(N_{{\rm ch}}\right)\right\rangle }{w\left(N_{{\rm ch}}\right)}=\label{eq:NDef}\\
 & =\frac{d\sigma_{J/\psi}\left(y,\,\eta,\,\sqrt{s},\,n\right)/dy}{d\sigma_{J/\psi}\left(y,\,\eta,\,\sqrt{s},\,\langle n\rangle=1\right)/dy}/\frac{d\sigma_{{\rm ch}}\left(\eta,\,\sqrt{s},\,Q^{2},\,n\right)/d\eta}{d\sigma_{{\rm ch}}\left(\eta,\,\sqrt{s},\,Q^{2},\,\langle n\rangle=1\right)/d\eta}\nonumber 
\end{align}
where $n=N_{{\rm ch}}/\langle N_{{\rm ch}}\rangle$ is the relative
enhancement of the charged particles in the bin, $w\left(N_{J/\psi}\right)/\left\langle w\left(N_{J/\psi}\right)\right\rangle $
and $w\left(N_{{\rm ch}}\right)/\left\langle w\left(N_{{\rm ch}}\right)\right\rangle $
are the self-normalized yields of $J/\psi$ and charged particles
(minimal bias) events in a given multiplicity class; $d\sigma_{J/\psi}(y,\,\sqrt{s},\,n)$
is the production cross-sections for $J/\psi$ with rapidity $y$
and $\langle N_{{\rm ch}}\rangle=\Delta\eta\,dN_{{\rm ch}}/d\eta$
charged particles in the pseudorapidity window $(\eta-\Delta\eta/2,\,\,\eta+\Delta\eta/2)$. 

\begin{figure}[ht]
\centering \leavevmode \includegraphics[height=4cm]{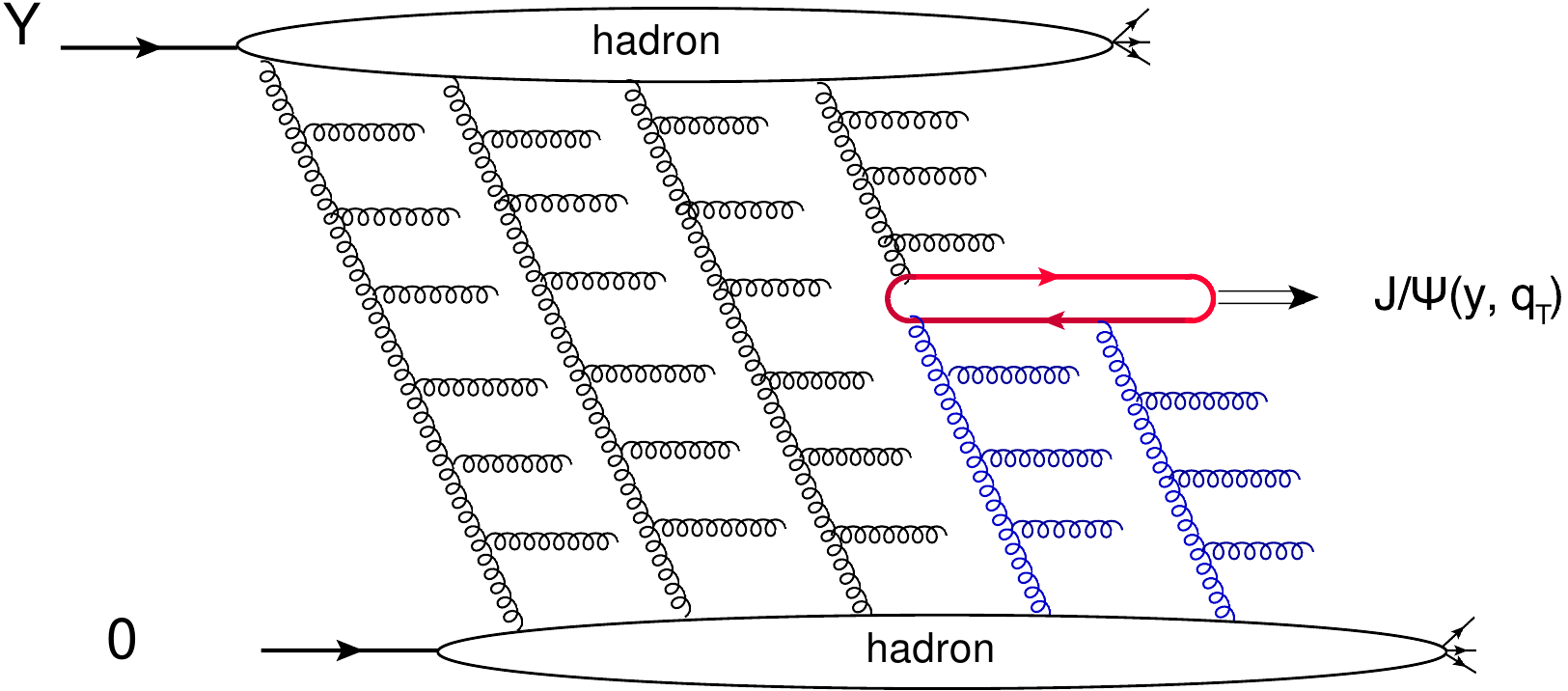}
\includegraphics[height=4cm]{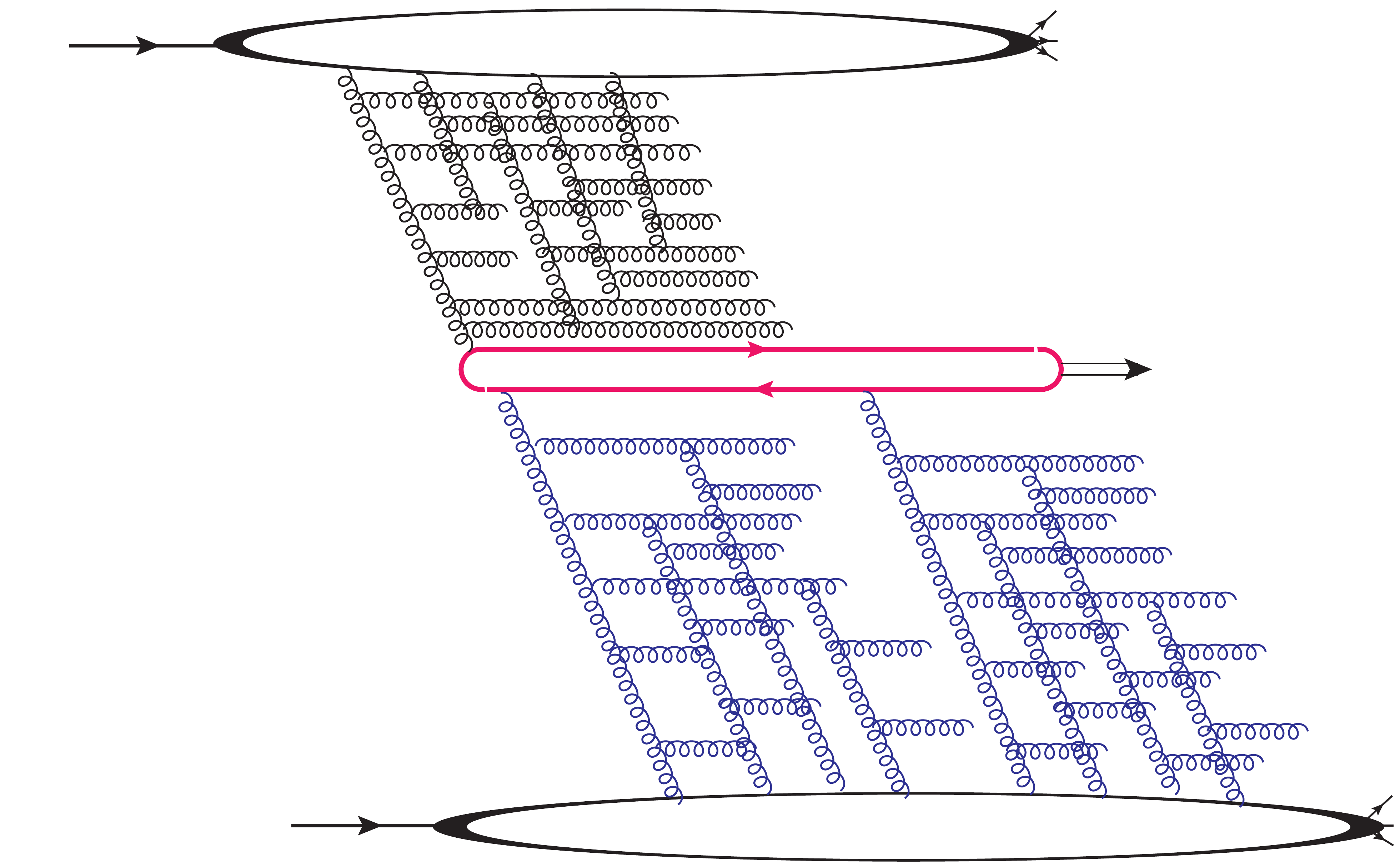}
\caption{Left: Contribution of the multigluon ladders to the enhanced multiplicity
of charged particles $N_{{\rm ch}}$ (\emph{traditional} mechanism)
which leads to linear dependence on $n$. Right panel: multiplicity
enhancement from BK cascade, subject of study in this paper.}
\label{gen2} 
\end{figure}

Traditionally it is assumed that enhanced multiplicity $N_{{\rm ch}}>\langle N_{{\rm ch}}\rangle$
might be due to contributions of mutligluon (multipomeron) configurations,
as shown in the Figure~\ref{gen2}. This mechanism leads to the multiplicity
dependence~
\begin{equation}
\frac{w\left(N_{J/\psi}\right)}{\left\langle w\left(N_{J/\psi}\right)\right\rangle }\,=\,\tilde{P}\left(N_{I\negmedspace P}\right)\,N_{I\negmedspace P}\,
\end{equation}
where $\tilde{P}\left(N_{I\negmedspace P}\right)$ is the probability
to produce $N_{I\negmedspace P}$ pomerons, and we expect that $N_{{\rm ch}}\sim N_{I\negmedspace P}$.
In case of inclusive charged particles production in a given multiplicity
class, we expect that the yield $w\left(N_{{\rm ch}}\right)/\left\langle w\left(N_{{\rm ch}}\right)\right\rangle $
is given by $\tilde{P}\left(N_{I\negmedspace P}\right),$ for this
reason the ratio~(\ref{eq:NDef}) evaluated with this mechanism should
have linear dependence on $n\sim N_{I\negmedspace P}/\langle N_{I\negmedspace P}\rangle$.
The experimental observation~\cite{PSIMULT,Alice:2012Mult} that
the $n$-dependence grows faster than linearly implies that other
mechanisms might give essential contribution. In the 3-pomeron mechanism
analyzed in this paper we expect that the multiplicity dependence
is enhanced due to a larger average number of particles produced from
each pomeron (see the right panel of the Figure~\ref{gen2}). We
expect that each such cascade (``pomeron'') should satisfy the nonlinear
Balitsky-Kovchegov equation. The increased multiplicity in individual
pomerons leads to modification of the dipole amplitude $N\left(y,\,\vec{r},\,\vec{b}\right)\to N\left(y,\,\vec{r},\,\vec{b},\,n\right)$.
To the best of our knowledge, currently there is no microscopic first-principle
evaluation of the dipole amplitude $N\left(y,\,\vec{r},\,\vec{b},\,n\right),$
for this reason we will use for it a phenomenological prescription.
As was demonstrated in~\cite{KOLEB,KLN,DKLN} the observed multiplicity
$dN_{{\rm ch}}/dy$ is roughly proportional to the saturation scale
$Q_{s}^{2}$. Since the distribution $dN_{{\rm ch}}/dy$ in the large
part of rapidity range has a very mild dependence on $y$, we may
assume that the saturation scale should grow in proportion to $n$~\cite{LERE,Lappi:2011gu},
\begin{equation}
Q_{s}^{2}\left(x,\,b;\,n\right)\,\,=\,\,n\,Q^{2}\left(x,\,b\right).\label{QSN}
\end{equation}
While at LHC energies it is expected that the typical values of saturation
scale $Q_{s}\left(x,\,b\right)$ fall into the range 0.5-1 ${\rm GeV}$,
in events with enhanced multiplicity this parameter might exceed the
values of heavy quark mass $m_{Q}$ and lead to interesting interplay
of large-$Q_{s}$ and large-$m_{Q}$ limits. In what follows we \emph{assume}
that there is no other modification of the dipole amplitude $N\left(y,\,\vec{r},\,\vec{b}\right)$
in this regime. Given the fact that dependence of $Q_{s}(x)$ on energy
is very mild, 
\begin{equation}
Q_{s}(x)\sim x^{-\lambda/2}\sim\left(\sqrt{s}/m_{M}\right)^{\lambda/2},\quad\lambda\approx0.2-0.3\,\,\,
\end{equation}
we see that the selection of high-multiplicity events is an effective
way to access the physics of the deep saturation regime, which otherwise
would require a significantly larger values of $\sqrt{s}$.

As we have argued in Ref.\cite{LESI}, for moderate values of $n$,
when the saturation scale is not very large, $Q_{s}\lesssim m_{Q}$,
the dominant contribution to the integrals over $r,\,r'$ in~(\ref{FD3})
comes from the vicinity of the saturation scale, $|r|\sim|r'|\sim Q_{s}^{-1}$,
where  the dipole amplitude  is given by~\cite{MUT} 
\begin{align}
N\,\, & =\,\,N_{0}\tau^{\bar{\gamma}}~~~~~~\mbox{with}~~~~~\bar{\gamma}=0.63,\label{VSQ}
\end{align}
For this reason in what follows we will also use the parametrization~(\ref{VSQ})
for semianalytical estimates of multiplicity dependence.

\begin{figure}
\includegraphics[width=14cm]{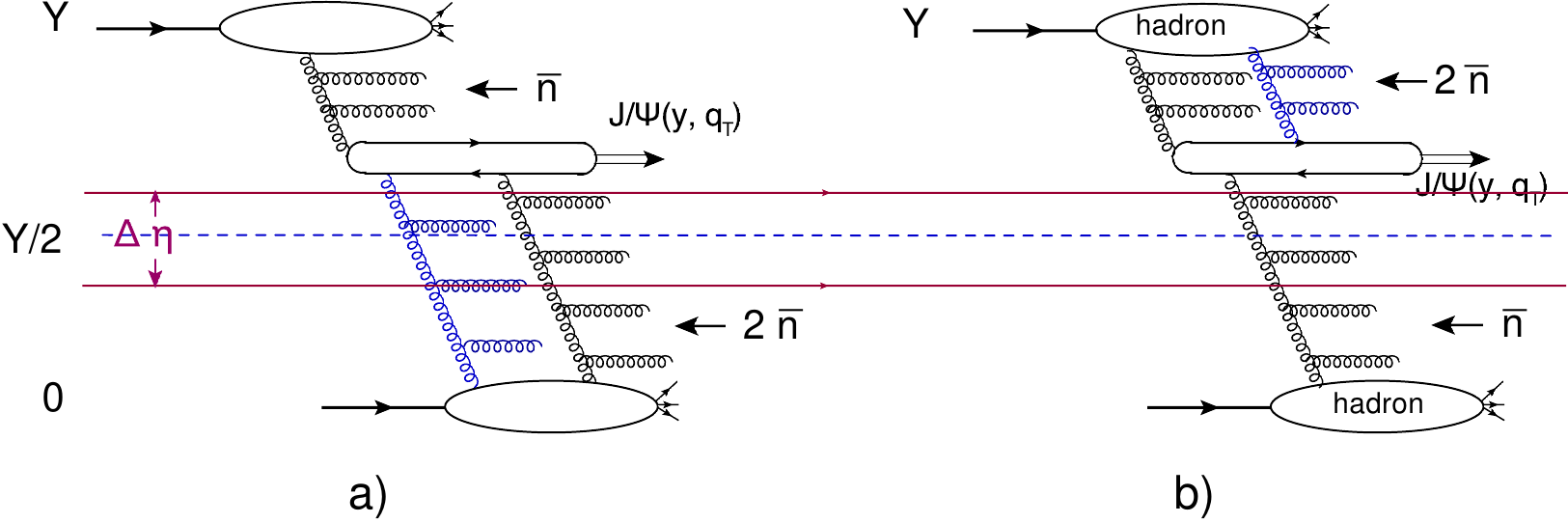}

\caption{Contribution of different configurations to the observed multiplicity
of charged particles when quarkonium ($J/\psi$) and charged particle
bins are well-separated by rapidity. We can clearly attribute observed
enhanced multiplicity (measured in the bin $\Delta\eta$) either to
1-pomeron or two-pomeron ladders.}
\label{DiagsMultiplicity1} 
\end{figure}

\begin{figure}
\includegraphics[width=14cm]{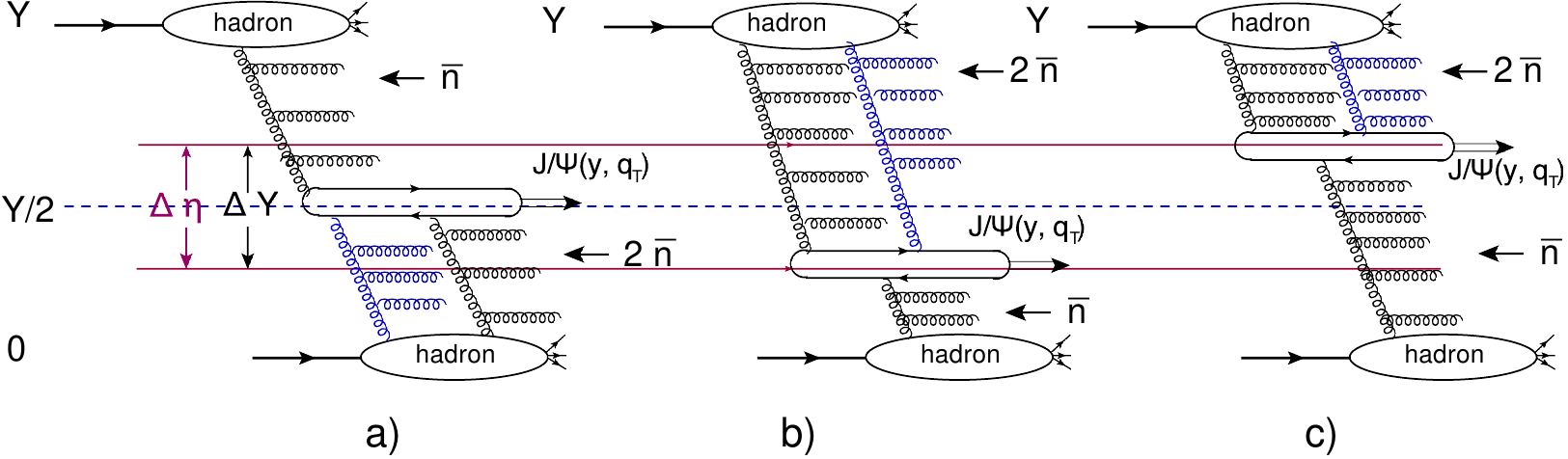}

\caption{In case when the quarkonia and charged particle bins overlap, the
observed ehanced multiplicity cannot be attributed either to single-pomeron
or to double-pomeorn yields, for this reason we have to integrate
over the rapidity of $J/\psi$ in the bin and sum over all possible
distributions between particles between the upper and lower pomerons,
such that $n_{1}+n_{2}+n_{3}=n$.}
\label{DiagsMultiplicity2} 
\end{figure}

In measurement of the multiplicity there are two different situations,
when the bins used for collecting charmonia and co-produced charged
particles are well-separated by rapidity, and when the bins overlap,
as shown in the Figures~\ref{DiagsMultiplicity1},~\ref{DiagsMultiplicity2}.
In the former case, we may unambiguously assign all the produced particles
either to one-gluon or to the two-gluon ladders in $t$-channel. Since
we assume that each gluon reggeizes independently and gives equal
contribution to observed average yields of charged particles, for
the two-pomeron case the enhanced multiplicity should be assigned
in equal parts to both pomerons. For this reason, the cross-section~(\ref{FD3})
for $n\not=1$ modifies as

\begin{eqnarray}
 &  & \frac{d\sigma\left(y,\,\sqrt{s},\,n\right)}{dy}\,\,=\frac{2C_{F}\mu_{F}}{\bar{\alpha}_{S}\,\pi}\,\int\frac{d^{2}Q_{T}}{(2\pi)^{2}}\,S_{h}^{2}\left(Q_{T}\right)\,\,\times\label{FD3-1}\\
 &  & \times\int_{0}^{1}dz\int_{0}^{1}dz'\int\frac{d^{2}r}{4\pi}\,\frac{d^{2}r'}{4\pi}\,\,\,\left\langle \Psi_{g}\left(r,\,z\right)\,\Psi_{J/\psi}\left(r,\,z\right)\right\rangle \,\left\langle \Psi_{g}\left(r',z'\right)\,\Psi_{J/\psi}\left(r',z'\right)\right\rangle \nonumber \\
 &  & \times\,\,\int dr"J_{1}\left(\mu_{F}\,r''\right)\Bigg\{ P(n)\nabla^{2}N\left(x_{1},\,r",\,n\right)\Delta N_{G}^{2}\left(x_{2},\,\vec{r},\,\vec{r}',\,1\right)\,\,\nonumber \\
 &  & +\,\,\int_{0}^{n}dn_{1}P\left(n_{1}\right)P\left(n-n_{1}\right)\,\nabla^{2}\,N\left(x_{2},\,r"\right)\,\Delta N_{G}\left(x_{1},\,\vec{r},\,\vec{r}',\,n_{1}\right)\Delta N_{G}\left(x_{1},\,\vec{r},\,\vec{r}',\,n-n_{1}\right)\Bigg\},\nonumber 
\end{eqnarray}
where we introduced shorthand notation
\begin{equation}
\Delta N_{G}\left(x,\,\vec{r},\,\vec{r}',\,n\right)\equiv N_{G}\left(x,\,\frac{\vec{r}+\vec{r}'}{2},\,n\right)\,-\,N_{G}\left(x,\,\frac{\vec{r}-\vec{r}'}{2},\,n\right),
\end{equation}
added explicitly the dependence on the relative multiplicity enhancement
parameter $n$, and for the sake of definiteness assumed that co-produced
charged particles are collected at higher rapidity $\eta$ than the
rapidity $y$ of $J/\psi$, as seen in the Diagram (A) of the Figure~\ref{DiagsMultiplicity1}
(the opposite case corresponds to inversion of sign of rapidity $y$). 

The evaluation of~(\ref{FD3-1}) requires knowledge of probability
distribution $P(n)$ for a pomeron to have relative enhancement of
multiplicity $n$. The theoretical evaluation of this quantity is
very involved. As was found experimentally~\cite{Abelev:2012rz},
the distribution of charged particles in $pp$ collisions for $n\gtrsim1$
is close to exponential behaviour $\sim\exp(-{\rm const}\,n)$ (modulo
logarithmic corrections)~\footnote{Such an exponential behaviour is an inherent feature of the Balitsky-Kovchegov
parton cascade as it has been demonstrated in~\cite{Kharzeev:2017qzs}}. However we would like to emphasize that this result cannot not be
applied to single pomerons because a convolution of any two parts
of pomeron does not satisfy the convolution indentity 
\begin{equation}
\sum_{N_{1}}\,P\left(N_{1},\,\left\langle N_{1}\right\rangle \right)P\left(N-N_{1},\,\left\langle N_{2}\right\rangle \right)=P\left(N,\,\left\langle N_{1}\right\rangle +\left\langle N_{2}\right\rangle \right).\label{eq:conv}
\end{equation}
On the other hand, it was predicted long ago~\cite{Levin:1993te}
that distribution of particles inside a single pomeron $P(N)$ might
be described by Poisson distribution. This distribution clearly satisfies~(\ref{eq:conv}).
 However, in order to reproduce the experimentally observed behaviour
in this approach, this would require resummation of infinite number
of pomerons with additional model assumptions for hadron-hadron collisions.
In view of this ambiguity, in what follows we will \emph{assume} that
the convolution of the distributions $P\left(n_{i}\right)$ in~(\ref{FD3-1})
just cancels in the ratio~(\ref{eq:NDef}), and in case when more
than one pomeron spans through the region covered by the detector,
the multiplicity is shared equally between pomerons, as we would get
with Poisson distributions. For this reason instead of the ratio~(\ref{eq:NDef})
we will work with a simplified ratio

\begin{align}
 & \frac{dN_{J/\psi}/dy}{\langle dN_{J/\psi}/dy\rangle}\,\,=\frac{d\tilde{\sigma}_{J/\psi}\left(y,\,\eta,\,Q^{2},\,n\right)/dy}{d\tilde{\sigma}_{J/\psi}\left(Y\,\eta,,\,Q^{2},\,\langle n\rangle=1\right)/dy}\label{eq:NDef-1}
\end{align}
where we use notation $d\tilde{\sigma}$ instead of $d\sigma$ for
the cross-section to emphasize that we took out normalization to probability
distribution of charged particles (factor $P(n)$). According to our
assumption, the integral over $n_{1}$ in~(\ref{FD3-1}) gets the
dominant contribution from the region when the enhanced multiplicity
is shared equally between pomerons, \emph{i.e}. $n_{1}=n/2$, so we
can rewrite the normalized cross-section as

\begin{eqnarray}
 &  & \frac{d\tilde{\sigma}\left(y,\,\sqrt{s},\,n\right)}{dy}\,\,=\frac{2C_{F}\mu_{F}}{\bar{\alpha}_{S}\,\pi}\,\int\frac{d^{2}Q_{T}}{(2\pi)^{2}}\,S_{h}^{2}\left(Q_{T}\right)\,\,\times\label{FD3-2}\\
 &  & \times\int_{0}^{1}dz\int_{0}^{1}dz'\int\frac{d^{2}r}{4\pi}\,\frac{d^{2}r'}{4\pi}\,\,\,\left\langle \Psi_{g}\left(r,\,z\right)\,\Psi_{J/\psi}\left(r,\,z\right)\right\rangle \,\left\langle \Psi_{g}\left(r',z'\right)\,\Psi_{J/\psi}\left(r',z'\right)\right\rangle \nonumber \\
 &  & \times\,\,\int dr"J_{1}\left(\mu_{F}\,r''\right)\Bigg\{\nabla^{2}N\left(x_{1},\,r",\,n\right)\Delta N_{G}^{2}\left(x_{2},\,\vec{r},\,\vec{r}',\,1\right)\,\,\nonumber \\
 &  & +\,\,\,\nabla^{2}\,N\left(x_{2},\,r"\right)\Delta N_{G}^{2}\left(x_{1},\,\vec{r},\,\vec{r}',\,\frac{n}{2}\right)\Bigg\}.\nonumber 
\end{eqnarray}
In the small-$n$ limit ($1\lesssim n\ll m_{Q}^{2}/Q_{s}^{2}(x)$)
we may use the dipole amplitude~(\ref{VSQ}) and get for the ratio~(\ref{eq:NDef})

\begin{figure}
\includegraphics[width=9cm]{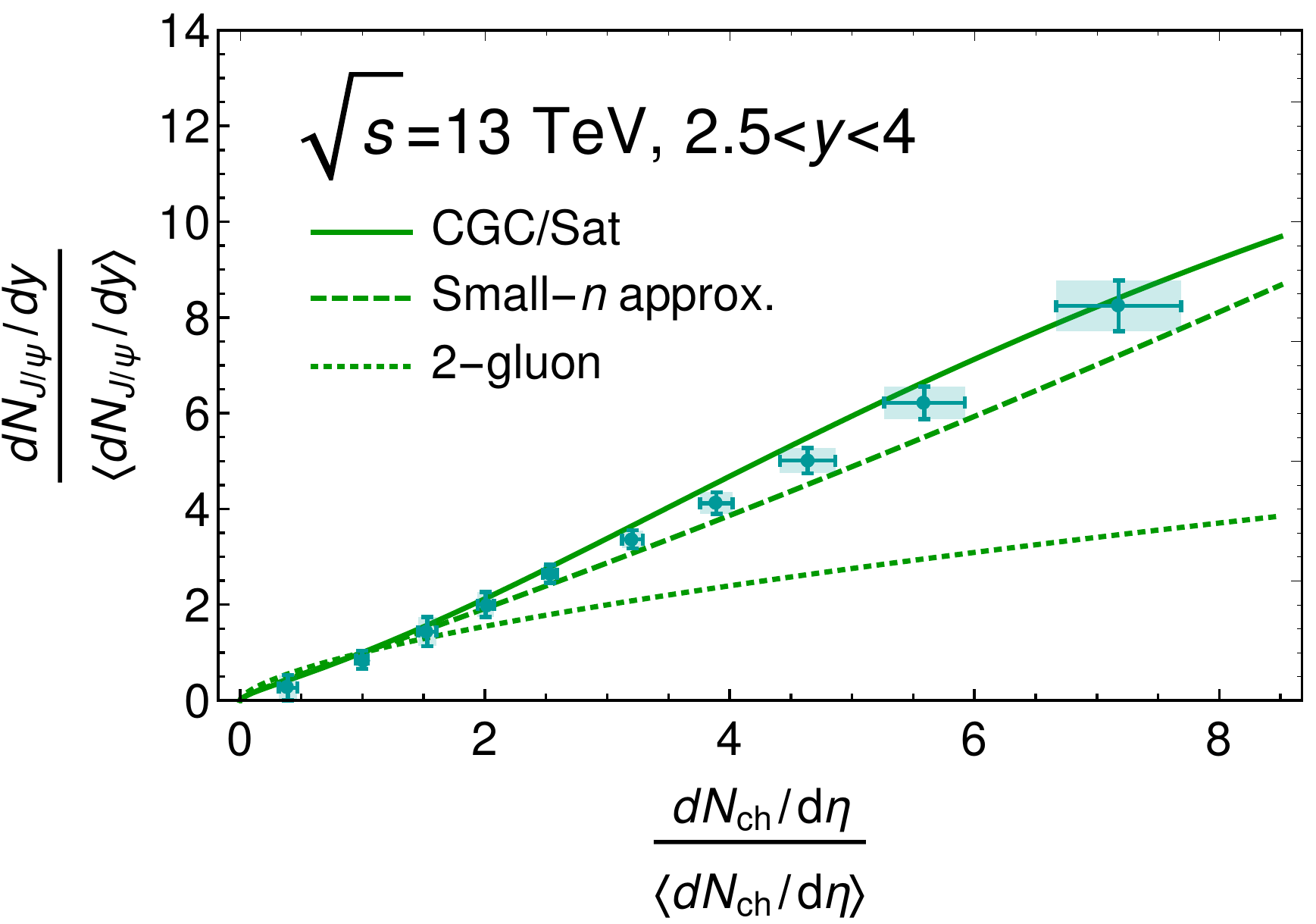}\includegraphics[width=9cm]{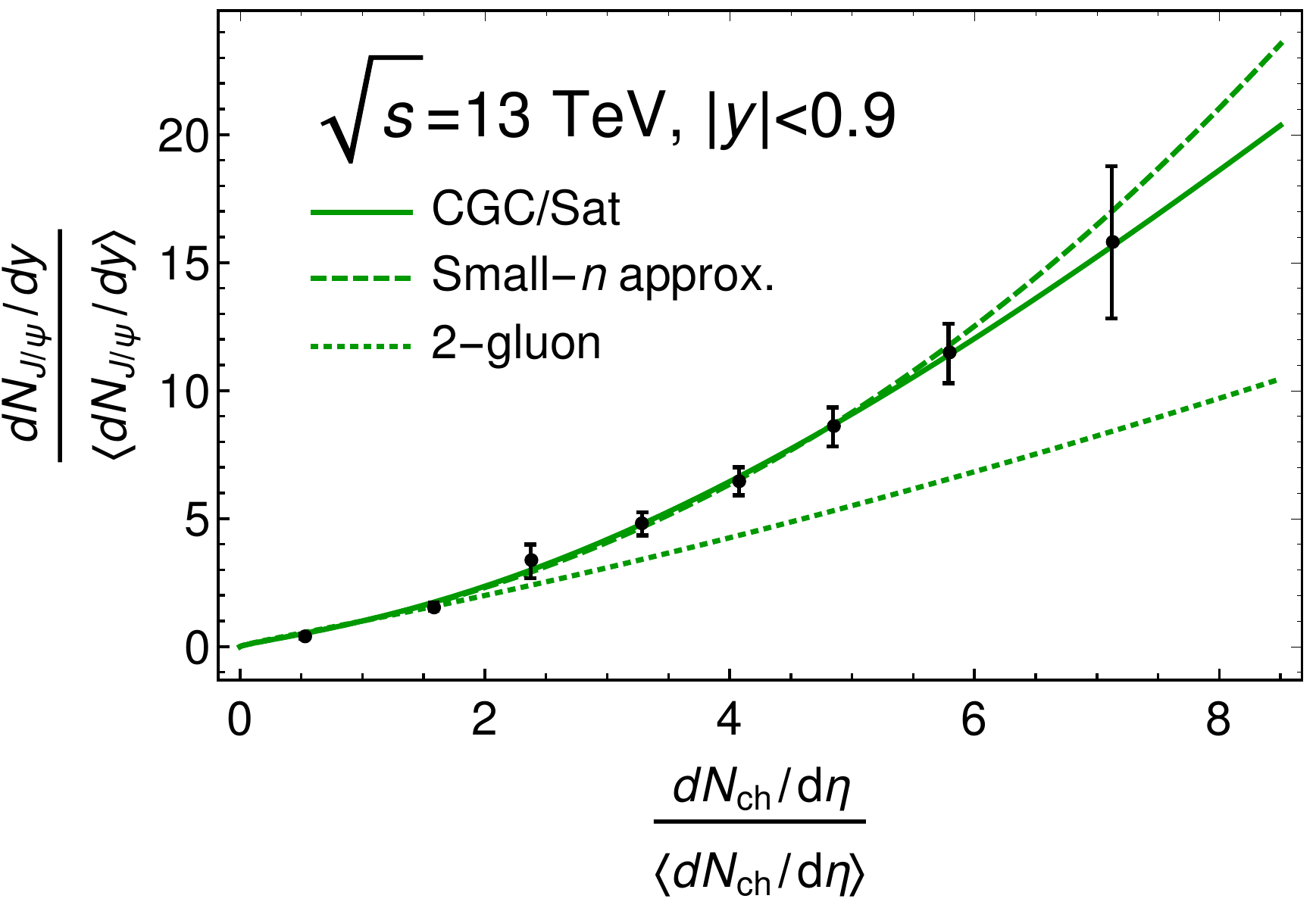}

\caption{Left: multiplicity dependence at forward rapidities ($2.5<y_{J/\psi}<4$),
charged particles are collected at central rapidities ($|\eta|<1$).
Experimental data are from~\cite{Khatun:2019slm}. Right: Evaluation
at central rapidities ($|y_{J/\psi}|<0.9$). Due to overlap of the
bins of charmonium and charged particles, we have an additional contribution
of the nonlinear term~\ref{FD4-1} which leads to more pronounced
$n$-dependence. Experimental data are from~\cite{PSIMULT,Alice:2012Mult}.
In both plots dashed curves labeled ``Small-$n$ approx.'' correspond
to evaluation with a dipole amplitude~(\ref{VSQ}). The curves labeled
``2-gluon'' in both plots correspond to $n$-dependence of the gluon-gluon
fusion mechanism, as explained in the text. In both plots we use the
LC Gauss wave function~\cite{Nemchik:1996cw} (see Appendix~\ref{sec:WFs})
for more details.}
\label{FwdCentral} 
\end{figure}

\begin{align}
\frac{dN_{J/\psi}/dy}{\langle dN_{J/\psi}/dy\rangle}\,\, & =\frac{Q_{s}^{4\bar{\gamma}}\left(x_{2}\right)Q_{s}^{2\bar{\gamma}}\left(x_{1}\right)\,n^{\bar{\gamma}}\,\,+Q_{s}^{4\bar{\gamma}}\left(x_{1}\right)Q_{s}^{2\bar{\gamma}}\left(x_{2}\right)\,\,\left(\frac{n}{2}\right)^{2\bar{\gamma}}}{Q_{s}^{4\bar{\gamma}}\left(x_{2}\right)Q_{s}^{2\bar{\gamma}}\left(x_{1}\right)+Q_{s}^{4\bar{\gamma}}\left(x_{1}\right)Q_{s}^{2\bar{\gamma}}\left(x_{2}\right)}\label{2PS1}\\
 & =\frac{1}{\kappa+(1/2)^{2\bar{\gamma}}}\Big\{\kappa n^{\bar{\gamma}}\,\,+\,\,\left(\frac{n}{2}\right)^{2\bar{\gamma}}\Big\}\nonumber 
\end{align}
where $x_{1,2}=m_{J/\psi}/\sqrt{s}\,\exp\left(\pm y\right)$ and $\kappa=Q_{s}^{2\bar{\gamma}}\left(x_{2}\right)/Q_{s}^{2\bar{\gamma}}\left(x_{1}\right)$
is a numerical coefficient. At the same time, for the gluon-gluon
fusion mechanism we expect that the corresponding $n$-dependence
would be simply $\sim n^{\bar{\gamma}}$ in this kinematics. As could
be seen from the left panel of the Figure~\ref{FwdCentral}, the
three-gluon mechanism~(\ref{FD3-1},\ref{eq:NDef}) describes the
experimental data~\cite{Khatun:2019slm} reasonably well, whereas
the 2-gluon contribution clearly underestimates the $n$-dependence.

More complicated is the situation when the $J/\psi$ and charged particle
bins partially overlap or coincide. In the overlap region we cannot
attribute enhanced multiplicity just to a single or double gluon ladder,
and instead have to average over all possible distributions of particles
$n_{1},\,n_{2},\,n_{3}$ among pomerons, such that $n_{1}+n_{2}+n_{3}=n$.
For this reason the cross-section for this case is given by (see Appendix~\ref{sec:Derivation}
for more details)

\begin{eqnarray}
 &  & \frac{d\sigma^{({\rm overlap})}\left(y,\,\sqrt{s},\,n\right)}{dy}\,\,=\frac{2C_{F}\mu_{F}}{\bar{\alpha}_{S}\,\pi}\,\int\frac{d^{2}Q_{T}}{(2\pi)^{2}}\,S_{h}^{2}\left(Q_{T}\right)\,\,\times\label{FD4}\\
 &  & \times\int_{0}^{1}dz\int_{0}^{1}dz'\int\frac{d^{2}r}{4\pi}\,\frac{d^{2}r'}{4\pi}\,\,\,\left\langle \Psi_{g}\left(r,z\right)\,\Psi_{J/\psi}\left(r,z\right)\right\rangle \,\left\langle \Psi_{g}\left(r',z'\right)\,\Psi_{J/\psi}\left(r',z'\right)\right\rangle \nonumber \\
 &  & \times\Delta\eta\int_{t_{{\rm min}}}^{t_{{\rm max}}}dt\,\int_{0}^{n_{1}+n_{2}\le n}dn_{1}dn_{2}\,P\left(\frac{n_{1}}{1-t}\right)P\left(\frac{n_{2}}{1-t}\right)P\left(\frac{n-n_{1}-n_{2}}{t}\right)\times\nonumber \\
 &  & \times\,\,\int dr"J_{1}\left(\mu_{F}\,r''\right)\Bigg\{\nabla^{2}N\left(x_{1},\,r",\,\frac{n-n_{1}-n_{2}}{t}\right)\Delta N_{G}\left(x_{2},\,\,\vec{r},\,\vec{r}',\,\frac{n_{1}}{1-t}\right)\,\Delta N_{G}\left(x_{2},\,\,\vec{r},\,\vec{r}',\,\frac{n_{2}}{1-t}\right)\,\,\nonumber \\
 &  & +\nabla^{2}N\left(x_{2},\,r",\,\frac{n-n_{1}-n_{2}}{t}\right)\Delta N_{G}\left(x_{1},\,\,\vec{r},\,\vec{r}',\,\frac{n_{1}}{1-t}\right)\Delta N_{G}\left(x_{1},\,\,\vec{r},\,\vec{r}',\,\frac{n_{2}}{1-t}\right)\Bigg\},\nonumber 
\end{eqnarray}
where we integrate over the rapidity of $J/\psi$ inside the bin using
the variable $t=(y-\eta_{{\rm min}})/\Delta\eta$, and over relative
multiplicities in each pomeron. In the case when the rapidity bins
used to collect charged particles and $J/\psi$ fully overlap ($y_{{\rm min}}=\eta_{{\rm min}}$,
$y_{{\rm max}}=\eta_{{\rm max}}$), the variable $t\in[0,\,1]$. As
we discussed earlier, currently we do not have a reliable first-principle
parametrization for $P(z)$, and for this reason in what follows we
will use a simplified \emph{assumption} that the full cross-section
might be approximated by a sum of three contributions, when the quarkonium
is produced at the center of the rapidity bin or at the borders, as
shown in the Figure~\ref{DiagsMultiplicity2}. The contribution of
the border region is given by~(\ref{FD3-2},\ref{2PS1}). In the
center of the bin, we may assume that the observed multiplicity $n$
is shared equally between the pomerons, so the contribution of this
region is given by

\begin{eqnarray}
 &  & \frac{d\tilde{\sigma}^{({\rm overlap,\,center})}\left(y,\,\sqrt{s},\,n\right)}{dy}\,\,\approx\frac{2C_{F}\mu_{F}}{\bar{\alpha}_{S}\,\pi}\,\int\frac{d^{2}Q_{T}}{(2\pi)^{2}}\,S_{h}^{2}\left(Q_{T}\right)\,\times\label{FD4-1}\\
 &  & \times\,\,\int_{0}^{1}dz\int_{0}^{1}dz'\int\frac{d^{2}r}{4\pi}\,\frac{d^{2}r'}{4\pi}\,\,\,\left\langle \Psi_{g}\left(r,z\right)\,\Psi_{J/\psi}\left(r,z\right)\right\rangle \,\left\langle \Psi_{g}\left(r',z'\right)\,\Psi_{J/\psi}\left(r',z'\right)\right\rangle \nonumber \\
 &  & \times\,\,\int dr"J_{1}\left(\mu_{F}\,r''\right)\Bigg\{\nabla^{2}N\left(x_{1},\,r",\,\frac{n}{3}\right)\Bigg(\,\Delta N_{G}\left(x_{2},\,\,\vec{r},\,\vec{r}',\,\frac{n}{3}\right)\Bigg)^{2}\,\nonumber \\
 &  & +\nabla^{2}N\left(x_{2},\,r",\,\frac{n}{3}\right)\Bigg(\,\Delta N_{G}\left(x_{1},\,\,\vec{r},\,\vec{r}',\,\frac{n}{3}\right)\Bigg)\Bigg\}.\nonumber 
\end{eqnarray}
With the dipole amplitude~(\ref{VSQ}) the expected large-$n$ dependence
of this additional term~(\ref{FD4-1}) is given by 
\begin{equation}
\frac{d\tilde{\sigma}^{({\rm overlap})}\left(y,\,\sqrt{s},\,n\right)}{dy}\sim n^{3\bar{\gamma}},\label{eq:n3}
\end{equation}
considerably more pronounced than~(\ref{2PS1}). In the experimentally
measured ratio~(\ref{eq:NDef}) for the cross-section $d\sigma$
we should take a weighted sum of the cross-sections~(\ref{FD3-2})
and (\ref{FD4-1}). We expect that with the parametrization~(\ref{VSQ})
the $n$-dependence should have a form
\begin{align}
\frac{dN_{J/\psi}/dy}{\langle dN_{J/\psi}/dy\rangle}\,\, & \approx\frac{1}{\kappa+(1/2)^{2\bar{\gamma}}+\Delta y\,(1/4)^{\bar{\gamma}}}\Big\{\kappa n^{\bar{\gamma}}\,\,+\,\,\left(\frac{n}{2}\right)^{2\bar{\gamma}}+\,\Delta\eta\,\left(\frac{n^{3}}{4}\right)^{\bar{\gamma}}\Big\}\label{2PS2}
\end{align}

\begin{figure}
\includegraphics[width=9cm]{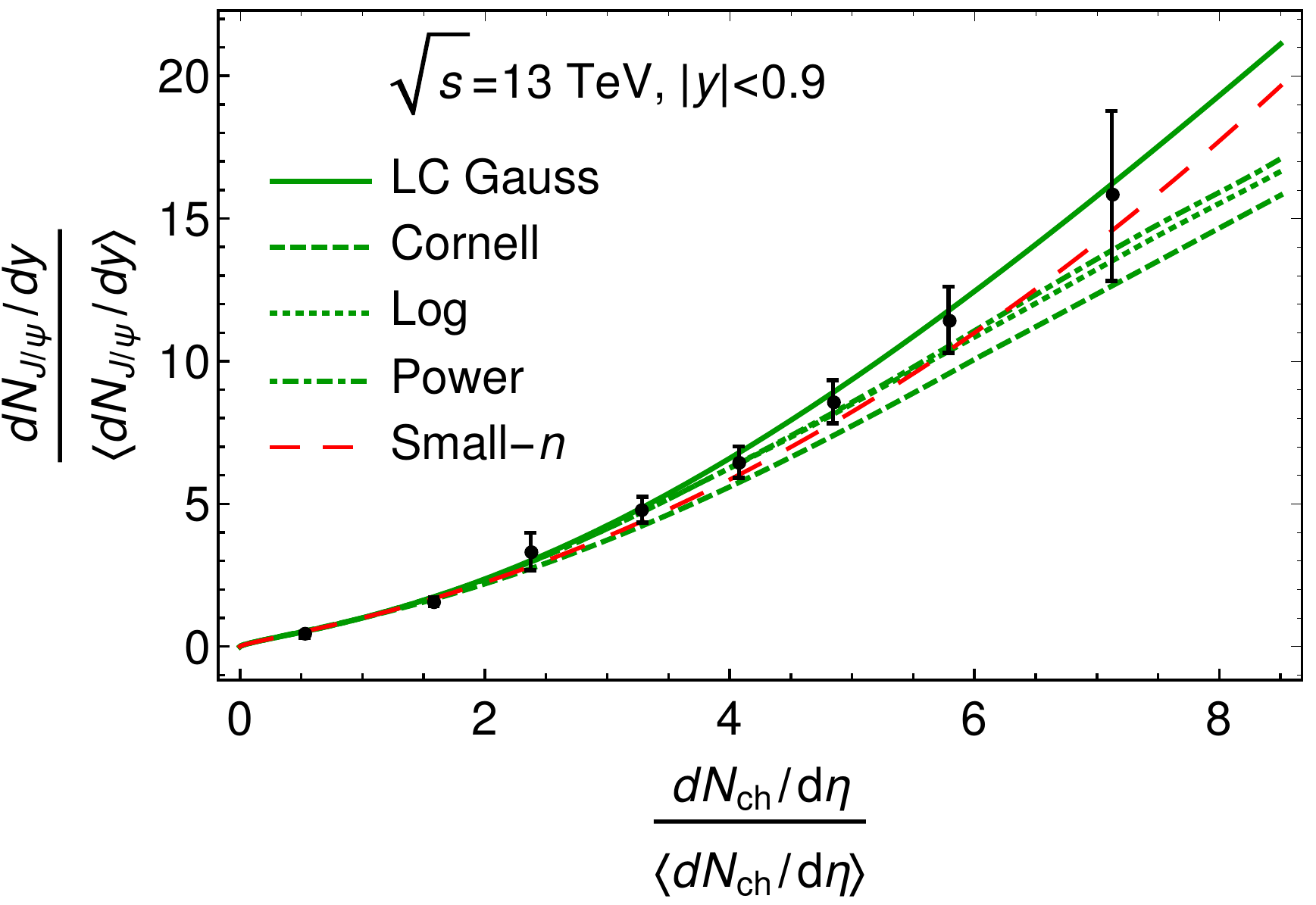}\includegraphics[width=9cm]{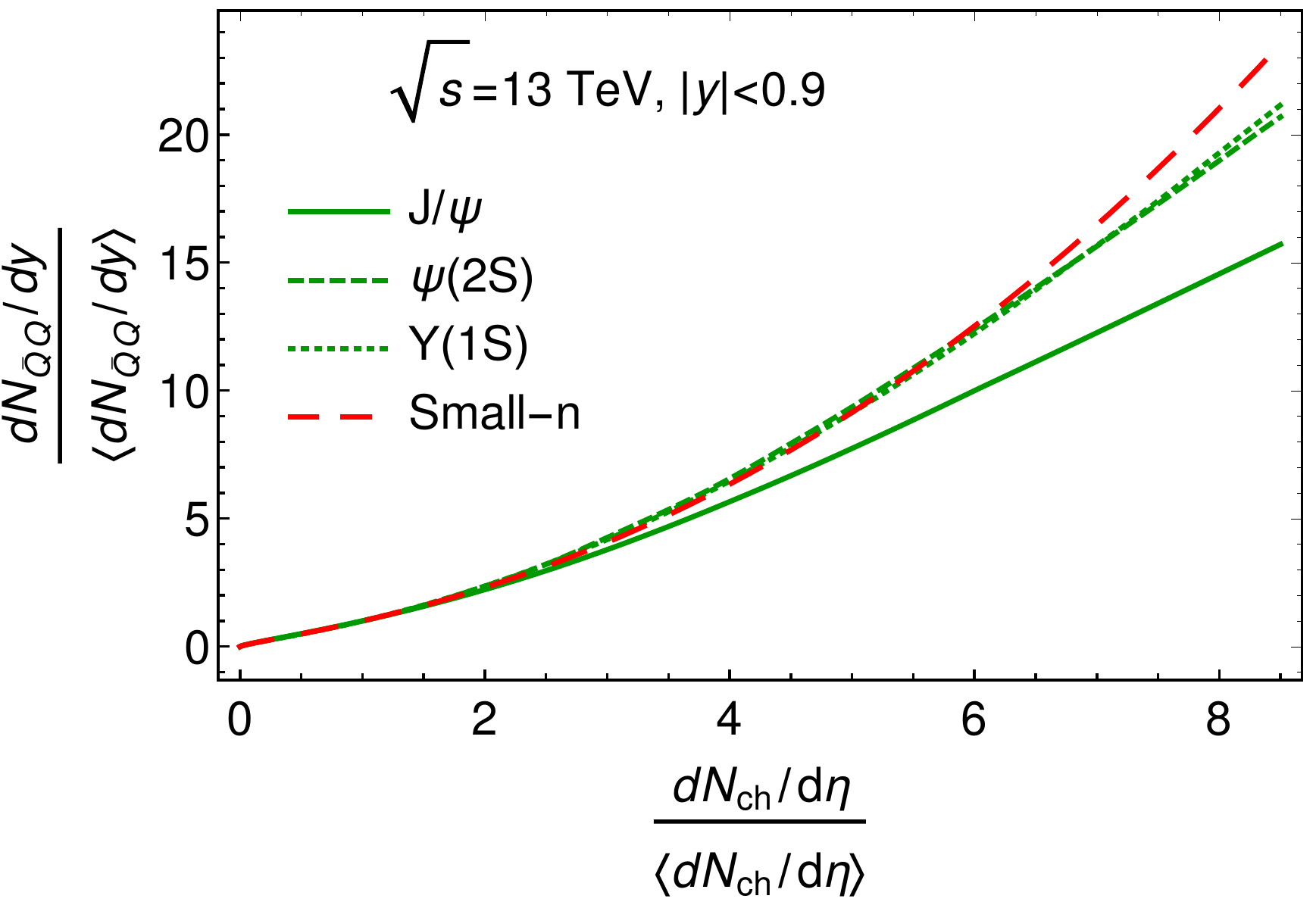}

\caption{Left: Dependence of the ratio~(\ref{eq:NDef}) on the choice of the
wave function. The solid line corresponds to evaluation with the light-cone
Gauss WF. The dashed and dotted lines correspond to evaluations with
boosted rest frame wave functions introduced in Appendix~\ref{sec:WFs}.
The experimental data are from~\cite{PSIMULT,Alice:2012Mult}. Right:
Comparison of $J/\psi$ results with predictions for different quarkonia
states, $\psi(2S)$ and $\Upsilon(1S)$ (we use Cornell WFs~\ref{sec:WFs}
instead of LC Gauss because it allows to make predictions for $J/\psi$
and higher excited states in the same approach. For other parametrizations
of the wave functions results are similar). }
\label{PsiALICE_WFs} 
\end{figure}

In the right panel of the Figure~\ref{FwdCentral} we've shown the
$n$-dependence with account of the additional term and compare it
with experimental data from~\cite{PSIMULT,Alice:2012Mult}. As we
can see, the model gives a reasonable description of experimental
data. In the Figure~\ref{PsiALICE_WFs} we show the dependence of
our predictions on the choice of the wave function, using for comparison
boosted rest frame wave functions described in Appendix~\ref{sec:WFs}.
As expected, this dependence is very mild. In the right panel of the
same Figure~\ref{PsiALICE_WFs} we show predictions for other quarkonia
states which might be within the reach of experimental searches in
the nearest future. We expect that both $\psi(2S)$ and $\Upsilon(1S)$
states might have slightly more pronounced $n$-dependence than the
$J/\psi$ meson.

\begin{figure}
\includegraphics[width=9cm]{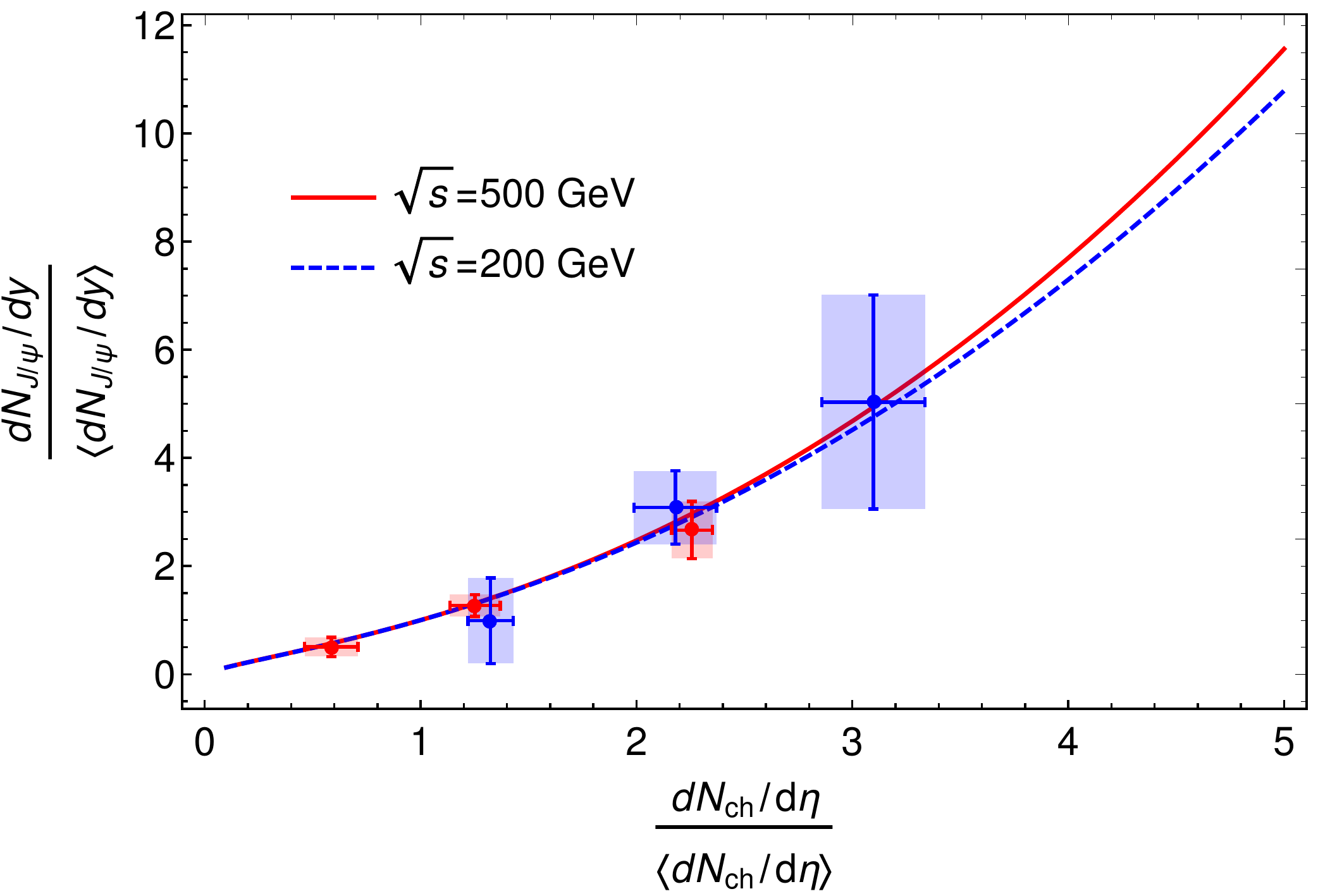}\includegraphics[width=9cm]{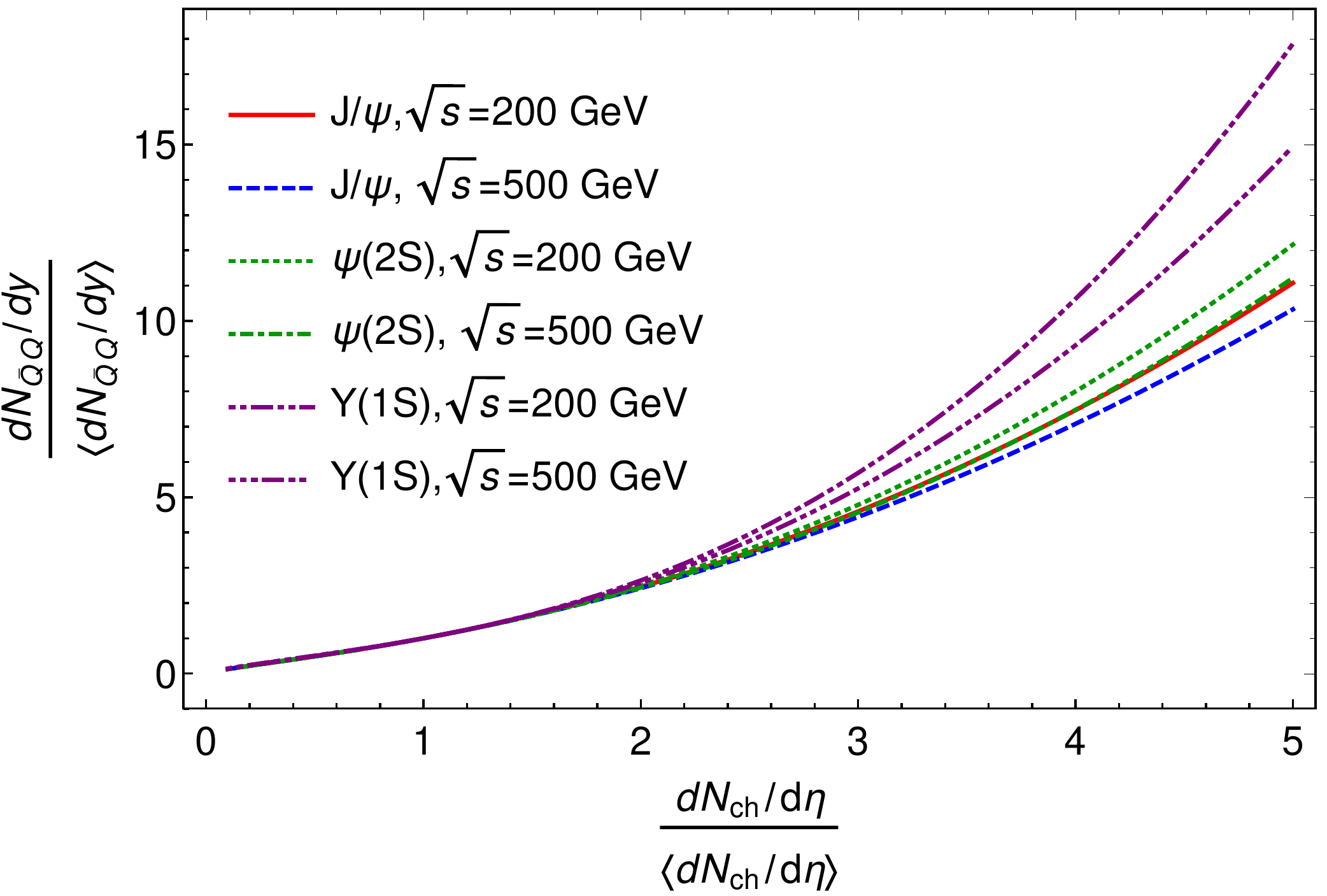}\caption{Left: Comparison of model results for the multiplicity with data from~\cite{Trzeciak:2015fgz,Ma:2016djk}.
Blue points and curves correspond to $\sqrt{s}=200$~GeV, red points
and curves correspond to $\sqrt{s}=500$~GeV. Right: Comparison of
$J/\psi$ results with predictions for $\psi(2S)$ and $\Upsilon(1S)$.}
\label{RHIC_QQ} 
\end{figure}

In the Figure~\ref{RHIC_QQ} we have shown results for $J/\psi$
multiplicity dependence in the RHIC kinematics, which are in a very
good agreement with experimental data~\cite{Trzeciak:2015fgz,Ma:2016djk}.
We also show predictions for $\Upsilon(1S)$ and $\psi(2S)$ which
are close to results for $J/\psi$. We hope that these predictions
will be checked very soon.

Finally, we would like to mention that the contribution of the last
term~(\ref{FD4}) might be singled out if the width of the pseudorapidity
bin $\Delta y$ used for collection of quarkonia is significantly
smaller than the width of the bin $\Delta\eta$ used for collection
of charged particles and includes it inside $(\Delta y\ll\Delta\eta),\,(y_{{\rm min}},\,y_{{\rm max}})\text{\ensuremath{\subset}}(\eta_{{\rm min}},\,\eta_{{\rm max}})$.
In this case the contribution of the border region is negligible,
and we expect that the $n$-dependence is given only by~(\ref{eq:n3}).
We hope in the future the experimentalists will be able to check this
prediction.

\section{Conclusions}

\label{sec:Concusion}In this paper, we analyzed the multiplicity
dependence of the two-gluon and three-gluon fusion mechanisms, assuming
that the observed multiplicity dependence is due to increase of the
average number of particles produced from each CGC pomeron. We found
that the two-gluon mechanism significantly underestimates the multiplicity
dependence, whereas the predictions of the three-gluon mechanism are
in reasonable agreement with available experimental data in a wide
energy range, from RHIC to LHC. We believe that this argument is a
strong evidence that the contribution of three-gluon mechanism might
be substantial in $J/\psi$ production. We call the experimentalists
to measure the multiplicity dependence of other charmonia (\emph{e.g}.
$\psi(2S)$ and $\Upsilon(1S)$), for which we expect approximately
the same multiplicity dependence as for $J/\psi$. Such measurement
would be an important argument in favor of 3-gluon mechanism in CGC
approach. We also predict that the suggested mechanism should provide
a strong $\sim n^{3}$ dependence in the kinematics when $J/\psi$
is produced at the center of the charged particle bin, i.e. the width
of the bin $\Delta y\ll\Delta\eta$.

Our predictions are sensitive to the large-distance behaviour of the
dipole amplitude, and going to higher values of $n$ potentially could
allow to test the saturation model implemented in a dipole amplitude.
However, experimental study of this regime is challenging due to a
rapid decrease of probability (yields) of such high-multiplicity events~~\cite{Abelev:2012rz},
which puts it out of reach of modern and forthcoming accelerators.

\begin{figure}[ht]
\centering \leavevmode \includegraphics[width=10cm]{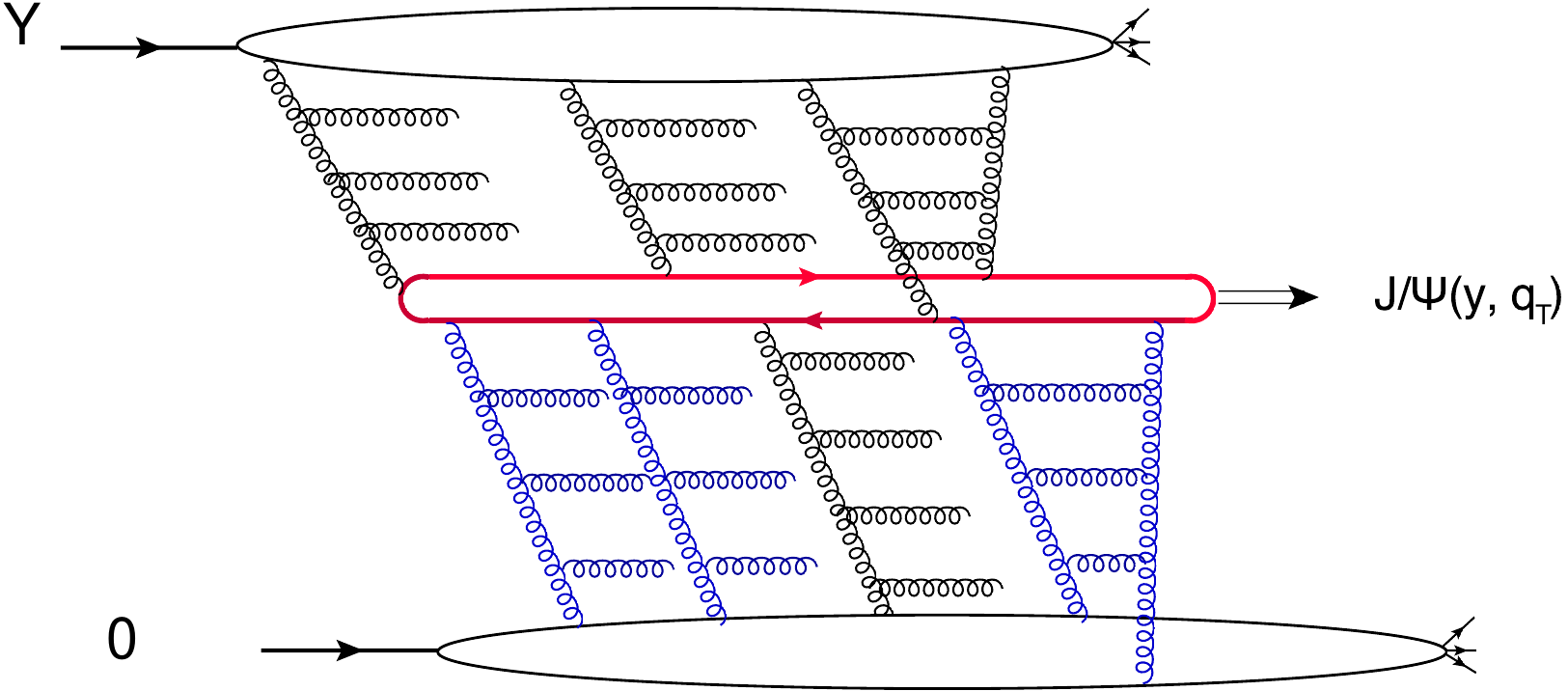}\caption{Alternative mechanism of multiplicity enhancement. The ladders in
this figure show the solution to the non-linear BK equation.}
\label{gen2-1} 
\end{figure}

The mechanism of multiplicity dependence suggested in this paper is
not unique: multiplicity might be also enhanced due to large number
of pomerons as shown in the the left panel of the Figure~\ref{gen2}
and in the Fig.~\ref{gen2}. We also should take into account that
each of the ladders (pomerons) shown in the Figure~\ref{gen2-1}
might lead to development of the Balitsky-Kovchegov cascade, as shown
in the right panel of the Figure~\ref{gen2}. We plan to address
these issues in a future publication~\cite{LS2020}.

\section{Acknowledgements}

We thank our colleagues at Tel Aviv university and UTFSM for encouraging
discussions. This research was supported by Proyecto Basal FB 0821(Chile),
Fondecyt (Chile) grants 1180118 and 1180232.

\appendix

\section{Derivation of the Eq.~(\ref{FD4})}

\label{sec:Derivation}~In this section we will evaluate the cross-section
of the pion production case when the bins used for collection of quarkonia
and charged particles overlap. In this setup we cannot attribute the
observed increase of multiplicity to any of the pomerons and have
to evaluate probability-weighted sum over all possible partitions
of the number of charged particles $N_{{\rm ch}}$ into three integers.
Since the quarkonium might be produced at any rapidity $y$ (not necessarily
at the center of the bin), this implies that the \emph{average} contribution
of each pomeron to enhanced multiplicity seen in the bin $\Delta\eta$
will also depend on $y$. In what follows it is convenient to work
with a variable $t=(y-\eta_{{\rm min}})/\Delta\eta$, which for the
case of the full overlap ($y_{{\rm min}}=\eta_{{\rm min}}$, $y_{{\rm max}}=\eta_{{\rm max}}$)
remains in the interval $t\in[0,\,1]$. 

In what follows we will use notations $N_{{\rm ch}}^{(1)},\,N_{{\rm ch}}^{(2)}$
and $N_{{\rm ch}}^{(3)}$ for the number of charged particles produced
from each of the pomerons, with 
\begin{equation}
N_{{\rm ch}}^{(1)}+N_{{\rm ch}}^{(2)}+N_{{\rm ch}}^{(3)}=N_{{\rm ch}}^{({\rm \Delta\eta})},\label{eq:Nch}
\end{equation}
where $N_{{\rm ch}}^{({\rm \Delta\eta})}$ is the total number of
charged particles produced inside the bin $\Delta\eta$. The probability
that a given pomeron of the length $\eta_{{\rm tot}}$ units of rapidity
produced $N_{{\rm ch}}^{({\rm tot})}$ particles is given by $P\left(N_{{\rm ch}}^{({\rm tot})}/\left\langle N_{{\rm ch}}^{({\rm tot})}\right\rangle \right)$,
where $\left\langle N_{{\rm ch}}^{({\rm tot})}\right\rangle $ is
the average number of particles produced by the whole pomeron. Since
the distribution $dN_{{\rm ch}}/d\eta$ almost does not depend on
rapidity, we may assume that inside each interval $\Delta\eta_{i}$
($i=1,2,3$) the distribution of produced charged particles is given
by the same function, $P\left(N_{{\rm ch}}^{(i)}/\left\langle N_{{\rm ch}}^{(i)}\right\rangle \right)$,
where the average number $\left\langle N_{{\rm ch}}^{(i)}\right\rangle =\left\langle N_{{\rm ch}}^{({\rm tot})}\right\rangle \Delta\eta_{i}/\eta_{{\rm tot}},$
and the length of the pomeron is given by
\begin{align}
\Delta\eta_{1} & =\Delta\eta_{2}=t\,\Delta\eta,\\
\Delta\eta_{3} & =(1-t)\Delta\eta.
\end{align}
The probability that 3 pomerons will share the particles according
to~(\ref{eq:Nch}) is given by
\begin{align}
 & P\left(N_{{\rm ch}}^{(1)},\,N_{{\rm ch}}^{(2)},\,N_{{\rm ch}}^{(3)}\right)=P\left(\frac{N_{{\rm ch}}^{(1)}}{\left\langle N_{{\rm ch}}^{(1)}\right\rangle }\right)P\left(\frac{N_{{\rm ch}}^{(2)}}{\left\langle N_{{\rm ch}}^{(2)}\right\rangle }\right)P\left(\frac{N_{{\rm ch}}^{(3)}}{\left\langle N_{{\rm ch}}^{(3)}\right\rangle }\right)=\\
 & =P\left(N_{{\rm ch}}^{(1)},\,\frac{N_{{\rm ch}}^{(1)}}{t\,\left\langle N_{{\rm ch}}^{(\Delta\eta)}\right\rangle }\right)P\left(\frac{N_{{\rm ch}}^{(2)}}{t\,\left\langle N_{{\rm ch}}^{(\Delta\eta)}\right\rangle }\right)P\left(\frac{N_{{\rm ch}}^{(3)}}{(1-t)\,\left\langle N_{{\rm ch}}^{(\Delta\eta)}\right\rangle }\right)=P\left(\frac{n_{1}}{t\,}\right)P\left(\frac{n_{2}}{t\,}\right)P\left(\frac{n-n_{1}-n_{2}}{1-t}\right),\nonumber 
\end{align}
where we introduced the variables $n_{i}=N_{{\rm ch}}^{(1)}/\left\langle N_{{\rm ch}}^{(\Delta\eta)}\right\rangle ,$
and $n=N_{{\rm ch}}^{(\Delta\eta)}/\left\langle N_{{\rm ch}}^{(\Delta\eta)}\right\rangle $.
The experimentally observable cross-section $d\sigma\left(M+N_{{\rm ch}}^{(\Delta\eta)}\right)$
to produce heavy quarkonium $M$ and $N_{{\rm ch}}^{(\Delta\eta)}$
particles in the bin $\Delta\eta$ might be related to the cross-sections
to produce quarkonium $M$ and $N_{{\rm ch}}^{(1)},\,N_{{\rm ch}}^{(2)},\,N_{{\rm ch}}^{(3)}$
charged particles in the rapidity interval $\Delta\eta$ from each
of the three pomerons as
\begin{equation}
d\sigma\left(J/\psi+N_{{\rm ch}}^{(\Delta\eta)}\right)=\sum_{N_{{\rm ch}}^{(1)},\,N_{{\rm ch}}^{(2)}}P\left(N_{{\rm ch}}^{(1)},\,N_{{\rm ch}}^{(2)},\,N_{{\rm ch}}^{(3)}\right)\,d\sigma\left(J/\psi+N_{{\rm ch}}^{(1)}+N_{{\rm ch}}^{(2)}+N_{{\rm ch}}^{(3)}\right).
\end{equation}
If the width of the bin $\Delta\eta$ is sufficiently large (so that
$\left\langle N_{{\rm ch}}^{(\Delta\eta)}\right\rangle \gg1$), we
may replace the summation over $N_{{\rm ch}}^{(1)},\,N_{{\rm ch}}^{(2)}$
with integral over $n_{1},\,n_{2}$ and obtain the Eq.~(\ref{FD4}).

\section{Wave functions of quarkonia}

\label{sec:WFs} In our evaluations, unless stated otherwise, we use
the so-called light-cone wave functions of $1S$ quarkonia states~\cite{Nemchik:1996cw}
\begin{equation}
\Phi_{{\rm LCG}}=\mathcal{N}\,z(1-z)\,\exp\left(-\frac{m_{Q}^{2}\mathcal{R}^{2}}{8\,z(1-z)}-\frac{2z(1-z)r^{2}}{\mathcal{R}^{2}}+\frac{m_{Q}^{2}\mathcal{R}^{2}}{2}\right),\label{eq:LCWF}
\end{equation}
where $\mathcal{N}$ and $\mathcal{R}$ are some numerical constants.
In the heavy quark mass limit, as well as in deeply saturated regime
we expect that the results should not depend on the choice of the
wave function at all, since only small dipole configurations in~(\ref{FD3})
give the dominant result. In the small-$r$ approximation we may replace
the wave function $\Phi(z,\,r)$ with its value in the point $r=0$,
which eventually cancels in the ratio~(\ref{eq:NDef}). However,
in order to assess the accuracy of this approximation, as well as
to have the wave functions of other quarkonia (primarily $\psi(2S)$
and $\Upsilon(1S)$ are of interest for experimental searches), we
also used the wave functions evaluated in the rest frame potential
models and boosted to the moving frame in the momentum space. Technically,
this prescription is given by a relation~\cite{Nemchik:1996cw,Hufner:2000jb}
\begin{align}
\Phi_{{\rm LC}}(z,\,r) & =\int d^{2}k_{\perp}e^{ik_{\perp}\cdot r_{\perp}}\left(\frac{k_{\perp}^{2}+m_{Q}^{2}}{2\,z^{3}(1-z)^{3}}\right)\psi_{{\rm RF}}\left(\sqrt{\frac{k_{\perp}^{2}+(1-2z)^{2}m_{Q}^{2}}{4z(1-z)}}\right),
\end{align}
where $\psi_{{\rm RF}}$ is the Fourier image of the eigenfunction
of the corresponding Schroedinger equation. For comparison we considered
three different choices of the potential: 
\begin{itemize}
\item The linearly growing Cornell potential 
\begin{equation}
V_{{\rm Cornell}}(r)=-\frac{\alpha}{r}+\sigma\,r,
\end{equation}
which is discussed in detail in~\cite{Eichten:1978tg,Eichten:1979ms} 
\item The potential with logarithmic large-$r$ behaviour suggested in~\cite{Quigg:1977dd},
\begin{equation}
V_{{\rm Log}}(r)=\alpha+\beta\,\ln r.
\end{equation}
\item The power-like potential introduced in~\cite{Martin:1980jx}, 
\begin{equation}
V_{{\rm pow}}(r)=a+b\,r^{\alpha}.
\end{equation}
\end{itemize}
We checked that the wave functions obtained with all three potentials
have similar shapes, and in case of $J/\psi$ are close to the light-cone
Gauss potential~(\ref{eq:LCWF}).


\begin{thebibliography}{10}
\bibitem{Maciula:2013wg}R.~Maciula and A.~Szczurek, Phys.~Rev.~D
\textbf{87}, no. 9, 094022 (2013) {[}arXiv:1301.3033 {[}hep-ph{]}{]}.

\bibitem{Chang:1979nn}C.~H.~Chang, Nucl.~Phys.~B \textbf{172},
425 (1980).

\bibitem{Baier:1981uk}R.~Baier and R.~Ruckl, Phys.~Lett.~\textbf{102B}
(1981) 364.

\bibitem{Berger:1980ni}E.~L.~Berger and D.~L.~Jones, Phys.~Rev.~D
\textbf{23}, 1521 (1981).

\bibitem{Bodwin:1994jh}G.~T.~Bodwin, E.~Braaten and G.~P.~Lepage,
Phys.~ Rev.~D \textbf{51}, 1125 (1995) Erratum: {[}Phys.~Rev.~D
\textbf{55}, 5853 (1997){]} {[}hep-ph/9407339{]}.

\bibitem{Maltoni:1997pt}F.~Maltoni, M.~L.~Mangano and A.~Petrelli,
Nucl.~Phys.~B \textbf{519}, 361 (1998) {[}hep-ph/9708349{]}.

\bibitem{Brambilla:2008zg}N.~Brambilla, E.~Mereghetti and A.~Vairo,
Phys.~Rev.~D \textbf{79}, 074002 (2009) Erratum: {[}Phys.~Rev.~D
\textbf{83}, 079904 (2011){]} {[}arXiv:0810.2259 {[}hep-ph{]}{]}.

\bibitem{Feng:2015cba}Y.~Feng, J.~P.~Lansberg and J.~X.~Wang,
Eur.~Phys.~J.~C \textbf{75}, no. 7, 313 (2015) {[}arXiv:1504.00317
{[}hep-ph{]}{]}.

\bibitem{Brambilla:2010cs} N.~Brambilla \textit{et al.}; 
 Eur. Phys. J. C\textbf{71}, 1534 (2011).

\bibitem{Baranov:2015laa} S.P.~Baranov, A.V.~Lipatov, N.P.~Zotov;
Eur. Phys. J. C\textbf{75}, 455 (2015).

\bibitem{Baranov:2016clx}S.~P.~Baranov and A.~V.~Lipatov, Phys.~Rev.~
D \textbf{96}, no. 3, 034019 (2017) {[}arXiv:1611.10141 {[}hep-ph{]}{]}.

\bibitem{ALICE:AAStrangeness}Abelev, B. \emph{et al}. (ALICE Collaboration),
Phys. Rev. Lett. \textbf{111}, 222301 (2013).

\bibitem{ALICE:AAStrangeness2}Abelev, B. \emph{et al}. (ALICE Collaboration),
Phys. Lett. B \textbf{728}, 216\textendash 227 (2014); Erratum: {[}\emph{ibid},
\textbf{734}, 409 (2014){]}.

\bibitem{ALICE:pAStrangeness}Abelev, B. \emph{et al}. (ALICE Collaboration),
Phys. Lett. B \textbf{728}, 25\textendash 38 (2014).

\bibitem{ALICE:pAStrangeness2}Adam, J. \emph{et al}. (ALICE Collaboration),
Phys. Lett. B \textbf{758}, 389\textendash 401 (2016).

\bibitem{ALICE:2017jyt}J.~Adam \emph{et al}. {[}ALICE Collaboration{]},
Nature Phys.~\textbf{13}, 535 (2017) {[}arXiv:1606.07424 {[}nucl-ex{]}{]}.

\bibitem{Thakur:2018dmp}D.~Thakur {[}ALICE Collaboration{]}, ``\emph{$J/\psi$
production as a function of charged-particle multiplicity with ALICE
at the LHC},'' arXiv:1811.01535 {[}hep-ex{]}.

\bibitem{Adam:2015ota}J.~Adam \emph{et al}. {[}ALICE Collaboration{]},
JHEP \textbf{1509}, 148 (2015) {[}arXiv:1505.00664 {[}nucl-ex{]}{]}.

\bibitem{Fischer:2016zzs}N.~Fischer and T.~Sjöstrand, JHEP \textbf{1701},
140 (2017) {[}arXiv:1610.09818 {[}hep-ph{]}{]}.

\bibitem{LESI} E.~Levin and M.~Siddikov, \textit{``$J/\psi$ production
in hadron scattering: three-pomeron contribution,''} Eur.\ Phys.\ J.\ C
\textbf{79} (2019) no.5, 376 doi:10.1140/epjc/s10052-019-6894-1 {[}arXiv:1812.06783
{[}hep-ph{]}{]}. 

\bibitem{PSIMULT} D.~Thakur {[}ALICE Collaboration{]}, \textit{``$J/\psi$
production as a function of charged-particle multiplicity with ALICE
at the LHC,''} arXiv:1811.01535 {[}hep-ex{]}. 
 

\bibitem{Alice:2012Mult} B. Abelev et al. {[}ALICE Collaboration{]},
\textit{``$J/\psi$ production as a function of charged particle
multiplicity in pp collisions at $\sqrt{s}$=7 TeV\textquotedbl{}},
Phys. Lett. B \textbf{712} (2012), 165.

\bibitem{KMRS} V.~A.~Khoze, A.~D.~Martin, M.~G.~Ryskin and
W.~J.~Stirling, \textit{``Inelastic $J/\psi$ and $\upsilon$ hadroproduction,''}
Eur.\ Phys.\ J.\ C \textbf{39}, 163 (2005), {[}hep-ph/0410020{]}.

\bibitem{MOSA} L.~Motyka and M.~Sadzikowski, \textit{``On relevance
of triple gluon fusion in $J/\psi$ hadroproduction,''} Eur.\ Phys.\ J.\ C
\textbf{75} (2015) no.5, {[}arXiv:1501.04915 {[}hep-ph{]}{]}. 
 

\bibitem{KHTU} D.~Kharzeev and K.~Tuchin, \textit{``Signatures
of the color glass condensate in J/psi production off nuclear targets,''}
Nucl.\ Phys.\ A \textbf{770} (2006) 40, {[}hep-ph/0510358{]};\,\,\,

\bibitem{KLNT} D.~Kharzeev, E.~Levin, M.~Nardi and K.~Tuchin,
Phys.\ Rev.\ Lett.\ \textbf{102} (2009) 152301, {[}arXiv:0808.2954
{[}hep-ph{]}{]}; \textit{``J/Psi production in heavy ion collisions
and gluon saturation,''} Nucl.\ Phys.\ A \textbf{826} (2009) 230,
{[}arXiv:0809.2933 {[}hep-ph{]}{]}; Nucl.\ Phys.\ A \textbf{924}
(2014) 47, {[}arXiv:1205.1554 {[}hep-ph{]}{]}.

\bibitem{KLN} D.~Kharzeev and M.~Nardi, \textit{``Hadron production
in nuclear collisions at RHIC and high density QCD,''} Phys.\ Lett.\ B
\textbf{507} (2001) 121;\,\, {[}nucl-th/0012025{]}. 
 D.~Kharzeev and E.~Levin, \textit{` `Manifestations of high density
QCD in the first RHIC data,''} Phys.\ Lett.\ B \textbf{523} (2001)
79, {[}nucl-th/0108006{]};\,\, D.~Kharzeev, E.~Levin and M.~Nardi,
\textit{``The Onset of classical QCD dynamics in relativistic heavy
ion collisions,''} Phys.\ Rev.\ C \textbf{71} (2005) 054903, {[}hep-ph/0111315{]};
\textit{``Hadron multiplicities at the LHC,''} J.\ Phys.\ G \textbf{35}
(2008) no.5, 054001.38 {[}arXiv:0707.0811 {[}hep-ph{]}{]}. 

\bibitem{DKLMT} F.~Dominguez, D.~E.~Kharzeev, E.~Levin, A.~H.~Mueller
and K.~Tuchin, \textit{``Gluon saturation effects on the color singlet
J/$\psi$ production in high energy dA and AA collisions,''} Phys.\ Lett.\ B
\textbf{710} (2012) 182, {[}arXiv:1109.1250 {[}hep-ph{]}{]}.

\bibitem{KLTPSI} D.~E.~Kharzeev, E.~M.~Levin and K.~Tuchin,
\textit{``Nuclear modification of the J/$\psi$ transverse momentum
distributions in high energy pA and AA collisions,''} Nucl.\ Phys.\ A
\textbf{924} (2014) 47 doi:10.1016/j.nuclphysa.2014.01.006 {[}arXiv:1205.1554
{[}hep-ph{]}{]}. 
 

\bibitem{KMV} Z.~B.~Kang, Y.~Q.~Ma and R.~Venugopalan, \textit{``Quarkonium
production in high energy proton-nucleus collisions: CGC meets NRQCD,''}
JHEP \textbf{1401} (2014) 056, {[}arXiv:1309.7337 {[}hep-ph{]}{]}.

\bibitem{GOLEPSI} E.~Gotsman and E.~Levin, \textit{``Energy evolution
of J/$\mathbf{\psi}$ production in DIS on nuclei,''} Phys.\ Rev.\ D
\textbf{98} (2018) no.3, 034014, {[}arXiv:1804.02561 {[}hep-ph{]}{]}.

\bibitem{KOLEB} \newblock Y.~V. Kovchegov and E.~Levin{\em {,
Quantum chromodynamics at high energy}} Vol.~33 (Cambridge University
Press, 2012). 

\bibitem{THOR} R.~S.~Thorne, \textit{``Gluon distributions and
fits using dipole cross-sections,''} AIP Conf.\ Proc.\ \textbf{792}
(2005) no.1, 324. 

\bibitem{BK} I.~Balitsky, \textit{``Operator expansion for high-energy
scattering\textquotedbl{}}, {[}arXiv:hep-ph/9509348{]};\,\, 
 \textit{``Factorization and high-energy effective action\textquotedbl{}},
\textit{Phys.\ Rev.} \textbf{D60}, 014020 (1999) {[}arXiv:hep-ph/9812311{]};\,\,\,\,
Y.~V.~Kovchegov, \textit{`` Small-x $F_{2}$ structure function
of a nucleus including multiple Pomeron exchanges\textquotedbl{}'}
\textit{Phys.\ Rev.} \textbf{D60}, 034008 (1999), {[}arXiv:hep-ph/9901281{]}.

\bibitem{GS1} J.~Bartels and E.~Levin, \textit{``Solutions to
the Gribov-Levin-Ryskin equation in the nonperturbative region,''}
Nucl.\ Phys.\ B \textbf{387} (1992) 617. doi:10.1016/0550-3213(92)90209-T

\bibitem{GS2} A.~M.~Stasto, K.~J.~Golec-Biernat and J.~Kwiecinski,
\textit{``Geometric scaling for the total gamma{*} p cross-section
in the low x region,''} Phys.\ Rev.\ Lett.\ \textbf{86} (2001)
596 {[}arXiv:hep-ph/0007192{]}.

\bibitem{GS3} E.~Iancu, K.~Itakura and L.~McLerran, \textit{``Geometric
scaling above the saturation scale,''} Nucl.\ Phys.\ A \textbf{708}
(2002) 327 {[}arXiv:hep-ph/0203137{]}. 
 

\bibitem{LETU} E.~Levin and K.~Tuchin, \textit{``Solution to the
evolution equation for high parton density QCD,''} Nucl.\ Phys.\ B
\textbf{573} (2000) 833, {[}hep-ph/9908317{]}. 

\bibitem{RESH} A.~H.~Rezaeian and I.~Schmidt, \textit{``Impact-parameter
dependent Color Glass Condensate dipole model and new combined HERA
data,''} Phys.\ Rev.\ D \textbf{88} (2013) 074016, {[}arXiv:1307.0825
{[}hep-ph{]}{]}.

\bibitem{DKLN} A.~Dumitru, D.~E.~Kharzeev, E.~M.~Levin and Y.~Nara,
\textit{`` ``Gluon Saturation in $pA$ Collisions at the LHC: KLN
Model Predictions For Hadron Multiplicities,''} Phys.\ Rev.\ C
\textbf{85} (2012) 044920 {[}arXiv:1111.3031 {[}hep-ph{]}{]}.

\bibitem{LERE} E.~Levin and A.~H.~Rezaeian, \textit{``Gluon saturation
and inclusive hadron production at LHC,''} Phys.\ Rev.\ D \textbf{82}
(2010) 014022, {[}arXiv:1005.0631 {[}hep-ph{]}{]}.

\bibitem{Lappi:2011gu}T.~Lappi, Eur.~Phys.~J.~C \textbf{71},
1699 (2011) {[}arXiv:1104.3725 {[}hep-ph{]}{]}.

\bibitem{MUT} A.~H.~Mueller and D.~N.~Triantafyllopoulos, \textit{``The
Energy dependence of the saturation momentum,''} Nucl.\ Phys.\ B
\textbf{640} (2002) 331 {[}hep-ph/0205167{]};\,\,

\bibitem{Abelev:2012rz}B.~Abelev \emph{et al}. {[}ALICE Collaboration{]},
Phys.~Lett.~B \textbf{712}, 165 (2012) {[}arXiv:1202.2816 {[}hep-ex{]}{]}.

\bibitem{LS2020}M. Siddikov \emph{et al}, \emph{in preparation}.

\bibitem{Khatun:2019slm}A.~Khatun {[}ALICE Collaboration{]}, arXiv:1906.09877
{[}hep-ex{]}.

\bibitem{Trzeciak:2015fgz}B.~Trzeciak {[}STAR Collaboration{]},
J.~Phys.~Conf.~Ser.~\textbf{668}, no. 1, 012093 (2016) {[}arXiv:1512.07398
{[}hep-ex{]}{]}.

\bibitem{Ma:2016djk}R.~Ma {[}STAR Collaboration{]}, Nucl.~Part.~Phys.~Proc.
\textbf{276-278}, 261 (2016) {[}arXiv:1509.06440 {[}nucl-ex{]}{]}.

\bibitem{Nemchik:1996cw} J.~Nemchik, N.N.~Nikolaev, E.~Predazzi,
B.G.~Zakharov; 
 Z. Phys. C\textbf{75}, 71 (1997).

\bibitem{Hufner:2000jb}J.~Hufner, Y.~P.~Ivanov, B.~Z.~Kopeliovich
and A.~V.~Tarasov, Phys.~Rev.~D \textbf{62}, 094022 (2000) {[}hep-ph/0007111{]}.

\bibitem{Eichten:1978tg}E.~Eichten, K.~Gottfried, T.~Kinoshita,
K.~D.~Lane and T.~M.~Yan, Phys.~Rev.~D \textbf{17} (1978) 3090
Erratum: {[}Phys.~Rev.~D \textbf{21} (1980) 313{]}.

\bibitem{Eichten:1979ms}E.~Eichten, K.~Gottfried, T.~Kinoshita,
K.~D.~Lane and T.~M.~Yan, Phys.~Rev.~D \textbf{21} (1980) 203.

\bibitem{Quigg:1977dd}C.~Quigg and J.~L.~Rosner, Phys.~ Lett.~
\textbf{71B} (1977) 153.

\bibitem{Martin:1980jx}A.~Martin, Phys.~Lett.~\textbf{93B} (1980)
338.

\bibitem{Kharzeev:2017qzs}D.~E.~Kharzeev and E.~M.~Levin, Phys.~Rev.~D
\textbf{95}, no. 11, 114008 (2017) {[}arXiv:1702.03489 {[}hep-ph{]}{]}.

\bibitem{Levin:1993te}E.~Levin, Phys.~Rev.~D \textbf{49}, 4469
(1994).
\end{thebibliography}
\end{document}